\documentclass[%
 reprint,
 amsmath,amssymb,
 superscriptaddress,
 aps,
]{revtex4-2}

\usepackage{graphicx}% Include figure files
\usepackage{dcolumn}% Align table columns on decimal point
\usepackage{bm}% bold math
\usepackage{color}
\usepackage{soul}
\usepackage{placeins}
\usepackage{makecell}
\usepackage{booktabs}
\usepackage{longtable}
\usepackage{rotating}
\begin{document}

 \title{Firm-level supply chains to minimize unemployment and economic losses in rapid decarbonization scenarios}

\author{Johannes Stangl}
%\thanks{T.R. and G.H. contributed equally.}
\affiliation{Complexity Science Hub Vienna, Josefstädter Strasse 39, A-1080 Vienna, Austria}

\author{András Borsos}%
%\thanks{T.R. and G.H. contributed equally.}
\affiliation{Complexity Science Hub Vienna, Josefstädter Strasse 39, A-1080 Vienna, Austria}
\affiliation{National Bank of Hungary, Szabadság tér 9, 1054 Budapest}

\author{Christian Diem}
\affiliation{Complexity Science Hub Vienna, Josefstädter Strasse 39, A-1080 Vienna, Austria}

\author{Tobias Reisch}
\affiliation{Complexity Science Hub Vienna, Josefstädter Strasse 39, A-1080 Vienna, Austria}

\author{Stefan Thurner}
\email{Corresponding author, e-mail: stefan.thurner@meduniwien.ac.at}
\affiliation{Complexity Science Hub Vienna, Josefstädter Strasse 39, A-1080 Vienna, Austria}
\affiliation{Medical University of Vienna, Spitalgasse 23, A-1090 Vienna, Austria}
\affiliation{Santa Fe Institute, Santa Fe, 1399 Hyde Park Rd, NM 75791, USA}
\affiliation{Supply Chain Intelligence Institute Austria, Josefstädter Strasse 39, A-1080 Vienna, Austria}

\date{\today}% It is always \today, today,

\begin{abstract}
Urgently needed carbon emissions reductions might lead to strict command-and-control decarbonization strategies with potentially negative economic consequences. Analysing the entire firm-level production network of a European economy, we have explored how the worst outcomes of such approaches can be avoided. We compared the systemic relevance of every firm in Hungary with its annual CO$_2$ emissions to identify optimal emission-reducing strategies with a minimum of additional unemployment and economic losses. Setting specific reduction targets, we studied various decarbonization scenarios and quantified their economic consequences. We determined that for an emissions reduction of 20\%, the most effective strategy leads to losses of about 2\% of jobs and 2\% of economic output. In contrast, a naive scenario targeting the largest emitters first results in 28\% job losses and 33\% output reduction for the same target. This demonstrates that it is possible to use firm-level production networks to design highly effective decarbonization strategies that practically preserve employment and economic output.

\end{abstract}

\maketitle

The rapid decarbonization of the economy is one of the defining challenges of our time. The Intergovernmental Panel on Climate Change (IPCC) concludes that `rapid and far-reaching transitions in energy, land, urban and infrastructure [...] and industrial systems' are required \cite{ipcc_global_2018} to prevent catastrophic climate change. In this paper we explore how such `rapid and far-reaching' changes in the economic system might play out. We demonstrate the necessity of studying the supply chain linkages between firms for a proper understanding the dynamics of decarbonization and its consequences. We show how adverse economic and therefore societal consequences might be mitigated by explicitly incorporating information about the firm-level production network in the process of transformation. We base the current study on various lines of previous work.

The interplay of climate and economic systems has been extensively studied using various types of Integrated Assessment Models (IAM) that can either be classified as cost-benefit models or as process-based models. Cost-benefit models such as the DICE \cite{nordhaus_revisiting_2017}, FUND \cite{tol_optimal_1997}, or PAGE \cite{hope_critical_2013} model are used to estimate the `social cost of carbon' (SCC), quantifying the aggregate economic impacts of climate change and calculating `optimal' levels of global warming and an associated policy response. Process-based models study future development pathways and analyse mitigation options. They typically involve a great level of detail regarding the representation of the economic system, including several economic sectors and energy technologies, as well as the climate system. The IPCC uses an ensemble of process-based IAMs, such as IMAGE \cite{stehfest_integrated_2014}, MESSAGEix \cite{huppmann_messageix_2019}, AIM/GCE \cite{fujimori_development_2014}, GCAM \cite{calvin_gcam_2019}, REMIND-MAgPIE \cite{popp_economic_2011}, and WITCH \cite{bosetti_what_2011} to calculate emission pathways and analyze mitigation options. These models allow for a variety of analyses, such as more fine-grained and regionalized assessments of climate impacts, the socio-economic evaluation of climate policies, and the rollout of renewable energy technologies. More recent model generations, notably the E3ME model \cite{mercure_environmental_2018}, consider the economy in disequilibrium or allow for a representation of heterogeneous economic agents, such as the agent-based DSK-model \cite{lamperti_faraway_2018}. 

While decarbonization is largely a process of structural change \cite{ciarli_modelling_2019}, none of the aforementioned models provide a detailed representation of an economy's production network on the firm-level where the economic reconfiguration actually takes place during the transition. An exception is the LAGOM model family \cite{haas_agents_2005,mandel_lagom_2009,wolf_multi-agent_2013} that represents the economy as a firm-level production network which allows for the explicit study of shock propagation and the identification of critical firms in the production network \cite{lamperti_towards_2019}. The initialization from data, however, again happens on the industry sector level, mainly for reasons of data availability. This is also the case for other IAMs since national statistics and corresponding input-output analysis tools are only available on the sectoral level for most countries. Data on firm-level production networks, as presented in this study, could in principle be used to initialize the `Lagom regiO' model presented in \cite{wolf_multi-agent_2013}, informing the empirical firm-size distribution within sectors and the initial supply relations between firms. Recent research shows that the heterogeneity of the firm-level production network is essential to understand realistic economic dynamics; in particular, the propagation of disruptions and shocks can be strongly underestimated when using sectoral instead of firm-level production networks data \cite{diem_estimating_2023}. 

Most literature on decarbonization strategies and mitigation pathways focuses on mid- to long-term technological change. The present study presents a complementary perspective (see SI section S1 for a discussion). We explicitly focus on the short-term to analyze the role of the firm-level production network in the propagation and amplification of economic shocks originating from rapid decarbonization. The focus on short time scales (simulation horizon is one year and we assume no substantial technological or structural change in the production network) allows us to employ a framework of economic shock propagation.

Shock propagation through supply chains and resulting economic impacts were studied in the context of natural disasters like hurricane Katrina \cite{hallegatte_adaptive_2008} or the Japanese earthquake in 2011 \cite{inoue_firm-level_2019,carvalho_supply_2021}. The propagation of economic shocks from climate impacts on global supply chains has been studied \cite{bierkandt_acclimatemodel_2014,wenz_acclimatemodel_2014} on the sectoral level and environmentally extended input-output analyses allow for the study of the flow of embedded greenhouse gas emissions and other pollutants through regional and global sectoral supply chains \cite{leontief_environmental_1970,wiedmann_examining_2007,tukker_exiopol_2013}. A sectoral production network perspective has also been used to study the possibility of targeted sectoral carbon taxes \cite{king_targeted_2018}. Studies on economic shock propagation often employ the concept of `systemic relevance' of economic actors, meaning that the default (or shock) of an economic actor may result in disruptions or losses in different parts of the economy.

To estimate the systemic relevance of individual firms, we adopt the idea of the `economic systemic risk index' (ESRI) \cite{diem_quantifying_2022}. ESRI estimates the total economic output loss suffered by {\em all} affected firms in a production network, if an initial set of firms stop production. We modify the output-weighted ESRI (OW-ESRI) to estimate total job loss due to firm closures, calling it the `employment-weighted' ESRI (EW-ESRI). We estimate short-term worker displacement due to firm closures, not permanent job losses. Note that we are not considering permanent job losses that occur if industries fail to transition, but estimate worker displacement in the short-term, e.g., as a consequence of a climate policy that results in a production stop of one or several companies; see Methods.

OW-ESRI, EW-ESRI, and CO$_2$ emissions allow us to identify large emitters with low risk of inducing job and output losses. We identify these firms as 'decarbonization leverage points' and expect their removal to result in significant emission reductions and minimal job and output losses. To test this hypothesis and investigate the role of the firm-level production network, we simulate decarbonization strategies for Hungary's national economy using value-added tax (VAT) data \cite{borsos_unfolding_2020}. Each strategy represents a heuristic for hypothetically closing firms to reduce emissions while maintaining employment and output. These strategies are hypothetical and serve to probe network responses to initial shocks. We discuss the remarkable result that rapid decarbonization can have highly varying costs to society. Strategies targeting leverage points achieve substantial emission reductions while effectively securing employment and output. Ignoring the systemic economic relevance of targeted firms can lead to extreme unemployment and output loss. Our analysis shows the importance of considering the firm-level production network for decarbonization modeling and policy making.

\begin{figure*}
\centering
\includegraphics[width=14.0cm]{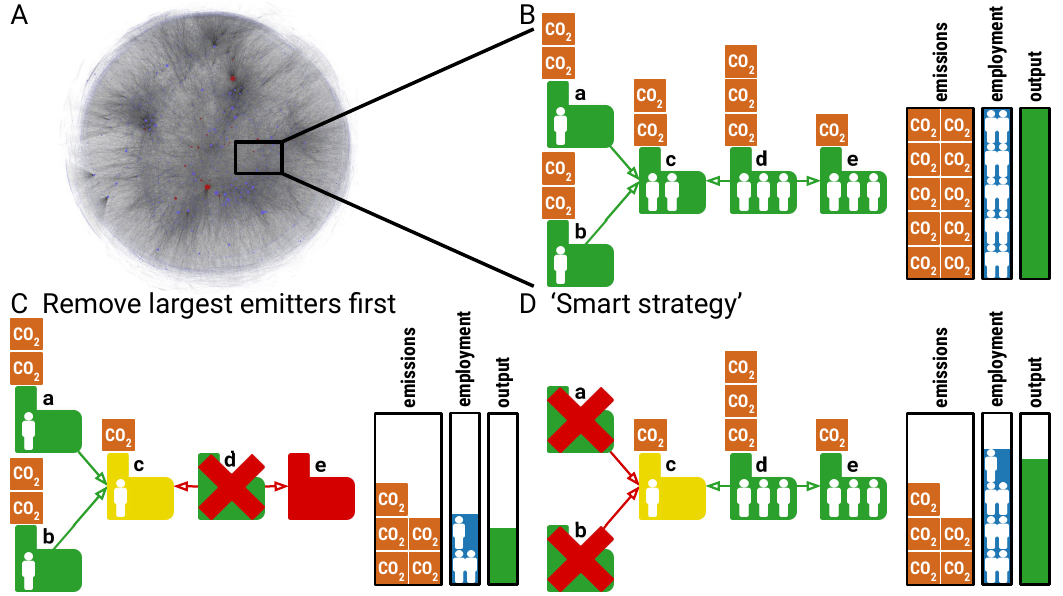}
\caption{Identifying decarbonization leverage points in firm-level production networks. (A) Production network of Hungary. (B) Schematic micro-level view of a production sub-network consisting of five firms. Every firm employs people, produces economic output and CO$_2$ emissions as a byproduct of its production. Total emissions, the number of jobs and total output of the sub-network are depicted in bars to the right. (C) `Remove largest emitters first' strategy. To effectively reduce emissions, firm \textbf{d} which is the largest emitter, is removed. Firm \textbf{e} loses its only supplier which causes it to stop its production and lay off its employees. Firm \textbf{c} loses one of its suppliers and reduces its production level by 50\%. In total, CO$_2$ emissions are reduced by 50\% while the total number of jobs is reduced by 70\%, similar to total economic output. 
(D) `Smart strategy' based on the identification of decarbonization leverage points. By closing  systemically irrelevant firms \textbf{a} and \textbf{b}, total CO$_2$ emissions are reduced by 50\% while the number of jobs drops by just 30\%. Total output is affected similarly. Since for this schematic figure a linear production function is assumed for all firms, only firm \textbf{c} is affected by the production stops of  \textbf{a} and \textbf{b}. The ratio of emission savings per job loss is maximal for this second strategy.}

\label{fig:toy_network}
\end{figure*}

{\bf Identifying decarbonization leverage points.}

Figure \ref{fig:toy_network} presents an overview of the framework for our analysis. Figure \ref{fig:toy_network}A depicts Hungary's 2019 production network. Nodes represent 243,399 Hungarian firms; links show 1,104,141 supply relations in Hungarian Forint (HUF). Node size reflects firm size and color denotes systemic risk. Figure \ref{fig:toy_network}B shows a micro-level view of the production network.
Five firms are connected through supply relations.
For simplicity, linear production functions are assumed for all firms and inputs in this conceptual plot. This means that firms adjust production levels to maximum available inputs upon a shock. Each firm employs people and emits CO$_2$ as a production byproduct. Total CO$_2$ emissions, employment, and economic output are depicted in the bars to the right. This is considered the initial 'normal' state of the network. To illustrate, imagine an unrealistic decarbonization strategy: a social planner aims for rapid decarbonization by closing the largest emitter. Shown in Fig. \ref{fig:toy_network}C: firm \textbf{d} experiences an imposed production stop. This results in the elimination of five units of CO$_2$ emissions in total, three from firm \textbf{d}, and one each from firms \textbf{c} and \textbf{e}. The cascading effect in the network leads to firm \textbf{e} losing its only supplier. Firm \textbf{e} stops production, laying off three employees. Firm \textbf{c} loses one supplier, thus maintaining lower production levels. It lays off half its employees, reducing CO$_2$ emissions by 50\% (one unit). In total, the removal of firm \textbf{d} decreases CO$_2$ emissions by 50\%, but jobs decrease by 70\% and total economic output declines accordingly.

Figure \ref{fig:toy_network}D depicts how considering the production network identifies decarbonization leverage points, minimizing job losses for emissions savings. We compute the employment-weighted ESRI to arrive at the 'emissions savings per job loss' (see Methods). Firms \textbf{a} and \textbf{b} are identified as decarbonization leverage points. Their removal causes firm \textbf{c} to lose two suppliers. It lays off half its employees, reducing CO$_2$ emissions by 50\%. The rest of the network remains unaffected. In total, removal of firms \textbf{a} and \textbf{b} decreases CO$_2$ emissions by 50\% but job losses only by 30\%. This means that relative emission savings per job loss are much higher than when targeting the largest emitter ignoring its importance in the network. Economic output remains at similarly high levels.

\section*{Results}

{\bf CO$_2$ emissions vs. EW-ESRI in the Hungarian economy.}

\begin{figure}[tb]
\centering
\includegraphics[width=8.0cm]
{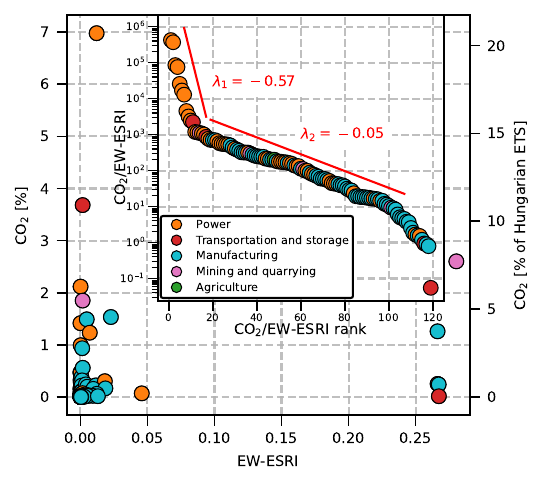}
\caption{CO2 emissions vs. EW-ESRI for Hungarian industrial firms
The y-axis on the left margin displays CO$_2$ emissions of firms in percentages of total Hungarian CO$_2$ equivalent emissions (CO$_2$ [\%]). The y-axis on the right margin shows CO$_2$ emissions as percentages of total Hungarian ETS emissions (CO$_2$ [\% of Hungarian ETS]). Colors indicate the NACE level 1 industry affiliation of every firm.
The inset shows the rank distribution of the emission savings per job loss on a semi-logarithmic scale. In particular, the percentages of total emissions divided by their respective EW-ESRI are shown. The y-axis spans seven orders of magnitude. Firms with CO$_2$/EW-ESRI $>$ 1000 are considered to be the primary decarbonization leverage points whose combined removal results in large total emissions savings and comparatively low levels of job and output loss.
}
\label{fig:CO2_vs_ESRI}
\end{figure}

Figure \ref{fig:CO2_vs_ESRI} shows CO$_2$ emissions of the 119 Hungarian firms that participate in the Emission Trading System (ETS) plotted against their systemic economic relevance as captured by EW-ESRI. The y-axis on the left margin captures the CO$_2$ emissions of firms as a percentage of total Hungarian CO$_2$ equivalent emissions (CO$_2$ [\%]). The y-axis on the right margin indicates CO$_2$ emissions as percentages of total Hungarian ETS CO$_2$ emissions (CO$_2$ [\% of Hungarian ETS]). The colors indicate the NACE level 1 industry affiliation for every firm. Most firms are classified as `C -  Manufacturing' (63)
and `D - Power’ (47). Most firms cluster around low emissions and low economic relevance (EW-ESRI). However, there are two distinct branches visible in Fig. \ref{fig:CO2_vs_ESRI}. One corresponds to firms with low emissions and comparatively high levels of EW-ESRI. The majority of these firms belong to the `Manufacturing' sector, which makes sense intuitively since many firms are dependent on the products from the manufacturing sector \cite{xu_inputoutput_2019}. The firm with the highest EW-ESRI of 0.28 is a firm operating in the `B - Mining and quarrying' sector. This means that approximately 28\% of jobs are lost, should this firm halt its production. 
This finding is reasonable since `Mining and quarrying' is one of the most central industrial sectors and encompasses activities at the very base of many economies \cite{xu_inputoutput_2019}.
The second branch corresponds to firms with high emissions and comparatively low EW-ESRI. These firms are candidates for potential decarbonization leverage points. Many firms from the `Power' sector are found in this branch. Note that there might be a bias emerging from the construction of the production network: since we only observe firm-firm supply relations but energy firms also serve many private customers, a proper accounting of the end consumers would make them more systemically important than we see them here. Another reason for the relatively low EW-ESRI values for power companies compared to manufacturing companies is their comparatively low market share within their respective sectors. In fact, the ESRI algorithm assumes that companies with a lower market share are more easily replaceable, suggesting that other energy firms could readily step in to cover any lost supply. See SI section S9 for more details.

The firm with the highest CO$_2$ emissions in 2019 accounts for 6.98\% of Hungary's total CO$_2$ emissions, but has an EW-ESRI of just 0.01. This means that only 1\% of jobs would be lost in the short-term if this firm would stop its production, but 7\% of annual CO$_2$ emissions could be saved. The fact that this firm operates in the `Power' sector underlines its potential replaceability in practice since a mix of renewable energy and storage capacities could make up for the missing fossil-based generation capacity. However, renewable energy firms require different skill sets than fossil fuel-based power plants and workers employed in fossil industries might be hesitant to transition towards green jobs, according to a recent study \cite{workers_green_energy_transition}. Additionally, even though many jobs will be created from the rollout of renewable energy sources, it may not be possible to fully replace the lost jobs in the short-term \cite{iea_world_2022}. 

The inset in Fig. \ref{fig:CO2_vs_ESRI} shows the rank distribution of the ratio of emissions savings per employment loss. In particular, it presents the CO$_2$ emissions of the individual firms measured in percentages of Hungary's total emissions, divided by their respective EW-ESRI. The y-axis is shown in logarithmic scale since the values of CO$_2$/EW-ESRI span seven orders of magnitude. There are 16 firms, mostly operating in the `Power' sector, whose ratio of CO$_2$/EW-ESRI exceeds 1000. Note the existence of two exponential regimes in the inset, characterized by two decay constants, $\lambda_1$ and $\lambda_2$. The exponential decay for CO$_2$/EW-ESRI $\sim e^{\lambda_i rank}$ for values above 1000 is found to be $\lambda_1 = -0.57$ which is about ten times larger than for values below 1000 and above ten, where we find $\lambda_2 = -0.05$. This means that the potential of the leading 16 firms to act as decarbonization leverage points is vastly larger than for the next 100 firms. Their relative annual emissions exceed their expected systemic risk with respect to employment loss by three orders of magnitude. \\

\begin{figure*}[p]
\centering
\includegraphics[width=12.8cm]{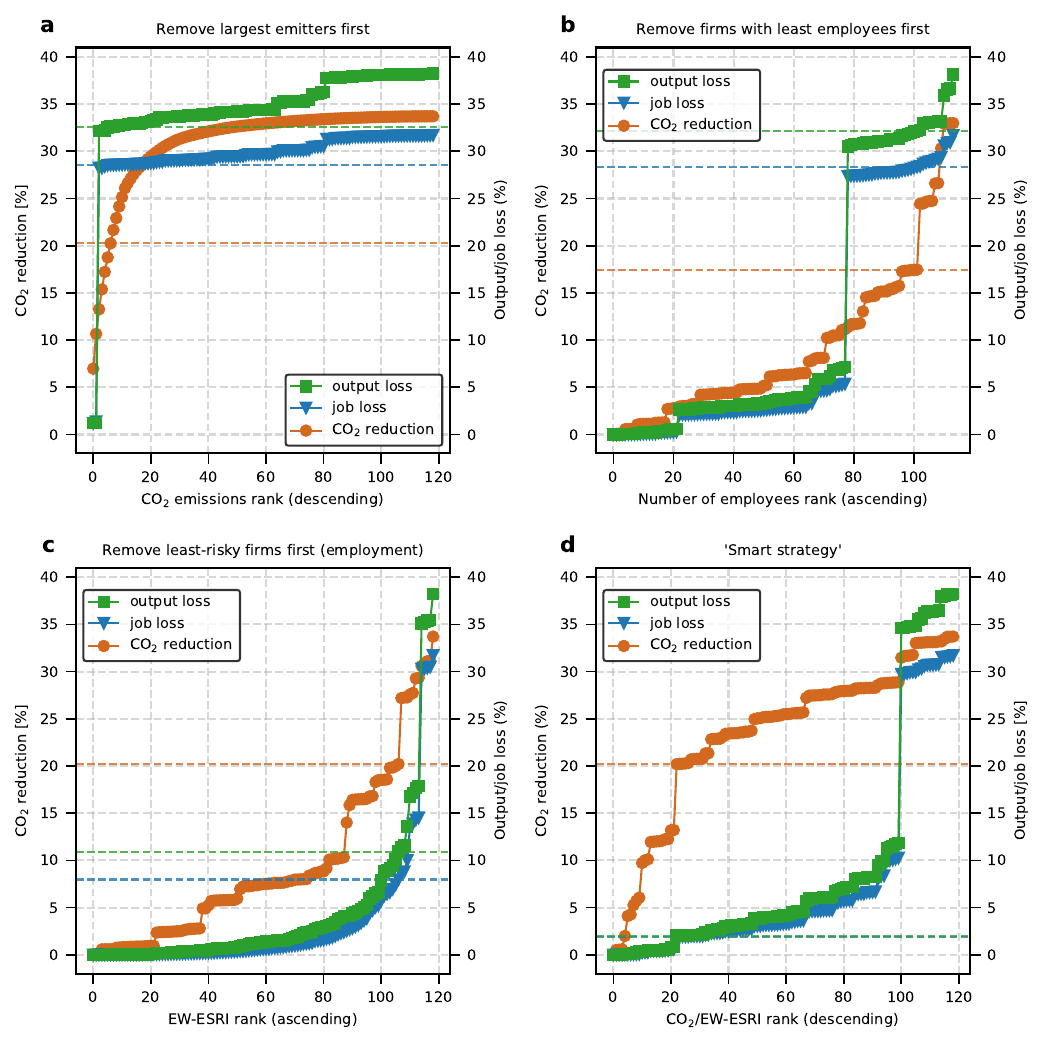}
\caption{Comparison of four decarbonization strategies
The total CO$_2$ reduction is indicated on the y-axis on the left margin; the estimated job and output loss are shown on the right margin. To compare different decarbonization strategies, we set an emission reduction benchmark of 20\% reduction of Hungary's total emissions (brown dashed line). The associated expected job and output loss from the removal of firms is indicated by a green and a blue dashed line, respectively (see Methods section for more details).
(A) Using the `Remove largest emitters first' strategy, the largest emitters are targeted first. 
By removing only 7 firms the emission benchmark is reached, however, the expected job and output loss immediately reach high values, since systemically highly relevant firms are removed early on.
(B) In the `Remove least-employees firms first' strategy, firms are closed according to their ascending numbers of employees. 
This results in only a gradual increase of expected job and output loss in the beginning, but fails to anticipate the effects of a systemically very important firm which triggers widespread job and output losses. 102 firms need to be closed in this strategy to reach the benchmark.
(C) In the `Remove least-risky firms first (employment)' strategy, firms are removed according to their ascending risk of triggering job loss, i.e., EW-ESRI; firms that are considered least systemically relevant for the production network are removed first. 
This leads to a very slowly increasing expected job and output loss, but 107 firms need to be closed to reach the benchmark. 
(D) In the `Smart strategy' firms are ranked according to their descending CO$_2$/EW-ESRI ratio; see inset of Fig. \ref{fig:CO2_vs_ESRI}. 
Emission reductions are achieved at comparatively low levels of expected job and output loss. 23 firms are closed to meet the target.}

\label{fig:cumulative_plots_strategies}
\end{figure*}

{\bf Comparison of  decarbonization strategies.} Figure \ref{fig:cumulative_plots_strategies} shows a comparison of four exemplary decarbonization strategies: The `Remove largest emitters first' strategy, which aims to reach the highest emission savings with a minimum number of firms to be removed from production, the 
`Remove least-employees firms first' strategy that aims at minimum job loss on the individual firm level,
`Remove least-risky firms first (employment)' strategy that tries to lose the least number of jobs per firm closed, including network effects, and the 
`Smart strategy' which aims at the largest possible emission reductions per job loss in the total economy.  
The different strategies each correspond to a different heuristic for the order in which the 119 Hungarian ETS firms are removed from the production network to reduce CO$_2$ emissions, as discussed in the Methods section. The corresponding results are seen in Fig. \ref{fig:cumulative_plots_strategies}
The x-axis of the different panels depicts the firm rank (firms are sorted) according to different criteria: (A) CO$_2$ emissions in descending order, (B) number of employees in ascending order, (C) EW-ESRI in ascending order, and (D) CO$_2$/EW-ESRI in descending order. The left y-axis margin states the saved relative CO$_2$ emissions by the cumulative removal of firms (brown). The y-axis on the right margin shows the expected job (blue) and output (green) loss from the cumulative removal of firms, as measured by the EW-ESRI and the OW-ESRI, respectively. 

Note that EW-ESRI and OW-ESRI enable us to calculate both, the direct job and economic output loss incurred by the removed firms, and the indirect losses experienced by all subsequently affected firms in the production network, for every decarbonization strategy, respectively. Taking the indirect effects in the production network into account leads to 3 to 42 times higher estimates of job loss and 6 to 23 times higher estimates of economic output loss for the decarbonization strategies depicted in Fig. \ref{fig:cumulative_plots_strategies}
This is discussed in the Methods section and in the SI section S3.

Saved emissions are given in percentages of Hungary's total emissions and add up to 33.7\% which is the total share of emissions of all ETS firms. The combined EW-ESRI and OW-ESRI of all firms being removed reach a maximum of 0.317 and 0.382 respectively. This means that approximately 31.7\% of jobs and 38.2\% of output would be lost, if {\em all} firms operating in the ETS would be removed from the production network. To compare different strategies with respect to their effectiveness in identifying firms, whose combined removal results in high emission savings and the lowest possible job and output loss, we introduce a specific benchmark (that we can set arbitrarily). For every strategy, we aim at a 20\% reduction of the country's total emissions by cumulatively removing firms according to the strategy-specific heuristic until the cumulative CO$_2$ emission savings add up to the closest value to 20\%. The resulting expected job and output loss for this benchmark are then compared for the four different decarbonization strategies. Note that the size of the area between the saved CO$_2$ emissions curve and the expected job and output loss curves is an indicator of the respective decarbonization strategy to effectively identify decarbonization leverage points. The benchmark results for all four decarbonization strategies are summarized in table \ref{tab:results_benchmark}. 

The four different decarbonization strategies shown in Fig. \ref{fig:cumulative_plots_strategies} display characteristic patterns for the saved CO$_2$ emissions and the expected job and output loss curves. 
For the `Remove largest emitters first' strategy in Fig. \ref{fig:cumulative_plots_strategies}A firms are ordered according to their descending CO$_2$ rank which is reflected in the cumulative emission savings curve (brown). 
It is clearly visible that closing the first firm already saves 7\% of emissions and that one needs to close 7 companies to reach the emissions reduction target of 20\%. The expected job loss curve (blue) and the expected output loss curve (green) show large jumps with the third firm being removed, followed by a slowly increasing regime before leveling off after the removal of 81 firms. We further read from the plot that to achieve the target of 20.25\% in emission reductions, approximately 32.61\% of output and 28.56\% of jobs are lost in this strategy.

The `Remove least-employees firms first' strategy that aims at minimizing job loss at each individual firm, shown in Fig. \ref{fig:cumulative_plots_strategies}B manages to keep expected job and output loss at low levels for the initially removed firms. But since this strategy focuses on job loss at the individual firm level, it fails to anticipate a highly systemically relevant firm whose closure results in high levels of expected job and output loss. Since CO$_2$ emissions are not explicitly considered in this strategy, emission savings only rise incrementally with additional firms with comparatively low numbers of employees being removed. To reduce CO$_2$ emissions by 17.35 \%, this strategy puts 32.24\% of output and 28.41\% of jobs at risk, while removing 102 firms from the production network. This strategy therefore fails to secure jobs and economic output, while delivering its emission savings.

The `Remove least-risky firms first (employment)' strategy that focuses on minimum expected job loss in the total economy per firm removed, as seen in Fig. \ref{fig:cumulative_plots_strategies}C, produces a cumulative emission savings curve that approximately increases linearly for the first 100 firms. The expected job loss increases slowly, as expected, since firms are ordered according to their ascending EW-ESRI rank. In order to achieve the goal of 20.21\% emission savings 107 firms need to be removed from the production network. As a consequence, 10.93\% of output and 7.97\% of jobs are lost. Note that this strategy represents a substantial advancement compared to the previous one, mainly because it takes network effects into account.

Finally, the `Smart strategy' is seen in Fig. \ref{fig:cumulative_plots_strategies}C, where one finds a steeply rising cumulative emissions saving curve similar to panel (A) and a slowly increasing job and output loss curve similar to panel (C). Only with the 98\textsuperscript{th} firm being removed, a large jump occurs. Using the CO$_2$/EW-ESRI ratio as the decarbonization heuristic, we are able to identify a ranking of firms whose combined removal results in high emissions savings and at the same time in low expected job and output loss. To achieve 20.19\% reduction in CO$_2$ emissions, the `Smart strategy' leads to a loss of 1.92\% of jobs and 2.02\% of output by removing 23 firms from the production network.

\begin{table*}[p]
\centering
\caption{Results for the 20\% CO$_2$ emission reduction target for the four decarbonization strategies presented in Fig. \ref{fig:cumulative_plots_strategies}.
The emission reduction target is reached at drastically different job losses, output losses, and number of firms removed from the production network.}

\begin{tabular}{lcccc}
\toprule
                                   strategy & \makecell{CO$_2$ reduction\\$\left[\%\right]$} & \makecell{job loss \\ (EW-ESRI) $\left[\%\right]$} & \makecell{output loss \\(OW-ESRI) $\left[\%\right]$} & \makecell{number of\\firms removed} \\

\midrule
              Remove largest emitters first &        20.25 &   28.56 &   32.61 &                       7 \\
         Remove least-employees firms first &        17.46 &   28.35 &   32.15 &                     102 \\
Remove least-risky firms first (employment) &        20.21 &    7.97 &   10.93 &                     107 \\
                             `Smart strategy' &        20.19 &    1.92 &    2.02 &                      23 \\
\bottomrule
\end{tabular}
\label{tab:results_benchmark}
\end{table*}

\FloatBarrier

\section*{Discussion}
The transition of the economy towards climate neutrality will create economic distress. Here, we present a framework to reduce negative side effects of the transition on employment and economic output by explicitly considering the firm-level production network of a national economy. 

We introduce EW-ESRI, a network-based measure, to estimate firms' systemic economic relevance in terms of employment. It captures firms' dependence on each other's production processes, indicating potential job loss if a firm faces distress or closure. We apply this measure to every firm in Hungary's economy using value-added tax (VAT) data to reconstruct the firm-level production network. We link this measure of systemic relevance to data on CO$_2$ emissions of Hungary's largest emitting firms and identify firms with high emissions and low systemic relevance as decarbonization leverage points. 

This allows us to simulate and compare different decarbonization strategies to understand EW-ESRI's usefulness in informing the design of decarbonization policies. The described scenarios demonstrate the importance of the firm-level supply network in decarbonization to limit economic costs. They illustrate the range of possible economic costs for the same emission reduction target under a command-and-control policy approach.

Prioritizing CO2 emissions over systemic importance results in a 28\% job loss and 33\% economic output reduction, politically and socially unacceptable. A strategy prioritizing jobs at each individual firm, ignoring effects on the production network, produces a similarly adverse outcome. Prioritizing systemic economic importance over CO2 emissions improves results, though removing many firms is necessary to reduce emissions significantly. Only a combined strategy considering both systemic importance and CO2 emissions achieves significant reductions with a small economic cost of 2\% job and economic output loss. Additional decarbonization strategy results are in SI section S1.

This means the same emission reductions can be achieved at substantially different job and economic output losses. A key insight from testing different decarbonization strategies is that some of the largest emitters often take up systemically important roles within an economy, but there are important exceptions. 

Differentiating risky from non-risky emitters requires studying their role in the production network. This result is especially relevant to be considered in climate-economic models such as IAMs. One avenue to do this is to integrate the presented firm-level production network approach with an existing macroeconomic model. Several models use an Input-Output Computable General Equilibrium (I-O CGE) approach to study decarbonization at the sectoral level. These models could be modified to allow the use of firm-level production networks instead of sector-level input-output tables, which, in principle, use the same mathematical formalism. Alternatively, modifying a suitable macroeconomic agent-based model could relax assumptions of I-O CGE models. Candidates include the E3ME \cite{mercure_environmental_2018} or DSK \cite{lamperti_faraway_2018} models, designed for firm-level dynamics but initialized from sectoral data.

The work faces several limitations. Conceptually, the framework does not replace other decarbonization modeling frameworks, like IAMs and integrating results obtained from this study into full-scale climate-economic models remains future work. 
The empirical part relies on a unique dataset of an entire economy's firm-level production network, but the dataset is incomplete with respect to imports, exports, and prices. Simplifying assumptions are necessary to assign production functions and estimate replaceability, detailed in \cite{diem_quantifying_2022}.
Prices are not considered as we adapt a network measure for short-term shock propagation in supply chains. Neglecting prices can be somewhat justified for the analysis of short-term economic dynamics, as several studies on shock-propagation in supply chains demonstrated \cite{hallegatte_adaptive_2008} \cite{inoue_firm-level_2019} \cite{inoue_firm-level_2019}. However, price dynamics becomes significant over longer time horizons, impacting shock propagation and network reconfiguration. Further research is needed to understand the role of price dynamics in shock propagation in firm-level production networks, extending simulation horizons.

In conclusion, our study emphasizes the importance of firm-level production networks in managing the economic outcomes of decarbonization. Considering the systematic importance of individual firms can substantially mitigate negative economic side effects of decarbonization strategies. Simultaneously considering systemic relevance and CO2 emissions could lead to new decarbonization policies, such as a supply-chain sensitive CO2 tax. Similar ideas were explored in financial networks \cite{poledna_elimination_2016} \cite{diem_what_2020} \cite{pichler_systemic_2021}. Future research should investigate the potential of such a taxation scheme to accelerate the restructuring of production networks towards decarbonization.

\newpage

\section*{Methods}

{\bf Data on the firm-level production network of Hungary.}
In Fig. \ref{fig:toy_network}A the network of all supply relations between Hungarian firms is seen. Nodes are firms and directed links represent actual supply relations between those firms, measured in monetary terms, i.e., Hungarian Forint (HUF). 243,399 firms and 1,104,141 supply relations are observed.
Node size corresponds to total strength $s_i$ which is defined as the sum of its in-strength and out-strength, $s_i = s_i^{in} + s_i^{out}$. The in-strength of a node $i$ represents the volume (monetary value) of purchased products and is defined as the sum of its weighted in-links, $s_i^{in} = \sum_{j=1}^n W_{ji}$. $W_{ij}$ represents the product volumes (monetary value) transacted from firm $i$ to firm $j$ in the year 2019. The out-strength of a node $i$ represents the sales of a firm and is defined as the sum of its weighted out-links, $s_i^{out} = \sum_{j=1}^n W_{ij}$. The total strength of a node is used as a proxy for firm size. For details regarding the reconstruction of the production network, see SI section S3. \\

{\bf The output-weighted economic systemic risk index (OW-ESRI).} Already a rough visual inspection of the Hungarian production network topology presented in Fig. \ref{fig:toy_network} suggests that different firms have different systemic relevance for the overall economy. The production stop of a firm with a high strength may affect a large part of the production network since many firms might lose significant fractions of their supply or demand. However, also the production stop of a smaller firm might affect the output of the production network considerably if it produces a good or service that serves as a {\em critical input} for other larger firms. The economic systemic risk index of firm $i$, ESRI$_i$, introduced by Diem et al. \cite{diem_quantifying_2022}, provides an estimate of the total output loss that is expected as the consequence of a production stop of firm $i$ in the short-term. By short-term we mean that the production network is kept constant. No rewiring occurs during the shock propagation and the adjustment of production levels. The ESRI is therefore a measure of a firm's systemic relevance in a given production network. This index is calculated as follows: the production network for a given year is initialized as $W(0)$ and each firm's production function is initialized to this initial state $t = 0$, defined by the inputs received by other firms. The firm of interest is removed from the production network at $t = 1$, which triggers a downstream supply shock to customers and an upstream demand shock to suppliers. The extent to which the shocks are propagated is dependent on the market share of the respective removed firm which is used as a proxy for its replaceability. It is assumed that firms with a high market share within their respective industries are harder to replace than firms with a low market share which determines the strength of the resulting downstream shock towards customers. The affected firms have to adapt their production levels according to their production function which triggers further downstream and upstream shocks. This process is repeated until a new equilibrium state is reached at $t = T$, for which production levels do no longer change significantly. To arrive at an assessment of overall economic output loss, the ESRI$_j$ of firm $j$ is calculated by weighting the remaining relative production levels $h_i(T)$ of all firms $i$ at time $T$ by their initial out-strength $s_i^{out}$. As this version of the ESRI accounts for economic output losses, we call it the output-weighted economic systemic risk index, OW-ESRI$_i$:

\begin{equation}
    \text{OW-ESRI}_j = \sum_{i=1}^n{\frac{s_i^{out}}{\sum_{l=1}^n {s_l^{out}}}} \left( 1 - h_i(T) \right) \, ,
\end{equation}
Note that final consumers are not considered in this study, due to a lack of data.\\

{\bf The employment-weighted economic systemic risk index (EW-ESRI).}
In order to additionally derive a measure of potential short-term job loss in the case firm $j$ being removed from the production network, we adapt the OW-ESRI by re-weighting the remaining relative production levels $h_i(T)$ of all firms $i$ at time $T$ by their initial number of employees $e_i$. We call this new measure the employment-weighted economic systemic risk index, EW-ESRI$_i$. We make the simplifying assumption that the needed labor of firm $i$ scales proportionally to its relative production level, $h_i(t)$. This results in the following equation
\begin{equation}
    \text{EW-ESRI}_j = \sum_{i=1}^n{\frac{e_i}{\sum_{l=1}^n{e_l}}} \left( 1 - h_i(T) \right) \, , 
\end{equation}
\noindent for which $e_i$ is the number of people employed at firm $i$. EW-ESRI yields the fraction of total short-term job loss in the given production network, should firm $i$ or several firms be removed from the network, i.e. stopping their production. For further details on the assumptions, the determination of production functions and other modeling that goes into the construction of ESRI, see \cite{diem_quantifying_2022}. Note that both the OW-ESRI and the EW-ESRI can also be calculated for a set of firms stopping their production.

Empirically, we are able to do this re-weighting of the ESRI for the Hungarian production network since a dataset on the number of employees of a large subset of firms is available to us through the Hungarian Central Bank. This dataset covers 2,333,975 jobs in total, which is 65.6\% of the 3,557,700 employees in the year 2019 according to official statistics \cite{hungarian_central_statistical_office_20225_nodate}.
In order to understand the coverage of our firm-level employment data set, we aggregate jobs by NACE level 1 category of their respective employing firms. The coverage of different economic sectors by our employment data varies but recovers essential features of the official employment distribution, such as the relative importance between sectors. In general, our data set exhibits the best coverage in the producing and in the service sectors and shows the worst coverage in the sectors higher up in the NACE classification scheme, like the public or education sector. A figure and a table of coverage and relative employment shares of each NACE level 1 sector for both official statistics and our aggregated employment data is included in the SI section S5.

{\bf Emissions of firms in the EU Emission Trading System (ETS).}
The Emission Trading System (ETS) is one of the cornerstones of the European Union's industrial climate policy. It requires industrial firms to acquire emission permits for their plants and installations either through free allocation or through trading extra permits with other firms. The ETS has undergone many design changes since its inception in 2005 which can be split into four phases. Since here we are dealing with data from 2019, phase three (2013 - 2020) is of main interest to us. In this phase, plants and installations from the following sectors are covered by the EU ETS: Power stations and other combustion plants $\geq$20MW; oil refineries; coke ovens; iron and steel plants; cement clinker; glass; lime; bricks; ceramics; pulp; paper and board; aluminum; petrochemicals; aviation; ammonia; nitric, adipic and glyoxylic acid products; CO$_2$ capture, transport pipelines and geological storage of CO$_2$ \cite{european_commission_eu_2015}. The European Union Transaction Log (EUTL) provides public access to data on the compliance of regulated installations, participants in the ETS, and transactions between participants \cite{european_commission_european_2022}. On top of the officially provided ETS data, a relational database is available to make the data more easily accessible \cite{jan_abrell_euetsinfo_2022}. This database was used in this study to make the necessary queries to access the Hungarian ETS emissions data and to aggregate verified emissions of installations to the Hungarian firms that own them. This resulted in a dataset of 122 firms and their verified emissions for the year 2019. This dataset was matched to the Hungarian firm-level production network dataset via the unique value-added tax (VAT) number of Hungarian firms. Three firms could not be matched which resulted in a final dataset of 119 firms for which both direct CO$_2$ emissions and other firm characteristics such as out-strength and number of employees are known. These 119 industrial firms emitted 21.72 million tonnes of CO$_2$ in the year 2019, which amounts to 33.7\% of Hungary’s total emissions of 64.44 million tonnes in 2019 \cite{international_energy_agency_hungary_2022}. The emissions of firms outside the Hungarian ETS remain unknown. This leads to an underestimation of saved CO$_2$ emissions for the discussed decarbonization strategies since indirect emission reduction effects cannot be considered. But since these 119 companies account for a third of Hungary's total emissions, this limited emissions dataset covers the essential CO$_2$ emitting firms of Hungary, which enables the key insights of our study. \\

{\bf Designing and assessing different decarbonization strategies.}
To empirically test our framework, we approximate hypothetical decarbonization efforts with the removal of firms from the Hungarian production network. A firm that is removed from the production network no longer supplies its customers nor does it place demand to its (former) suppliers in the subsequent time step. It also stops emitting CO$_2$. This hypothetical scenario allows us to quantify the worst-case outcomes in terms of job and economic output loss of a strict command-and-control approach towards decarbonization. In our simulation, a decarbonization strategy is realized as follows. We first rank firms according to four different charachteristics, CO$_2$ emissions, number of employees, systemic importance, and CO$_2$ emissions per systemic importance. Then, for every of these four strategies (shown in Fig. \ref{fig:cumulative_plots_strategies}) firms are cumulatively removed from the production network to assess the effects of the given heuristic. The first data point in Fig. \ref{fig:cumulative_plots_strategies} represents the highest ranked firm being removed from the production network. The second data point represents the highest and the second highest ranked firm, according to the respective heuristic, and so forth until all ETS firms are removed from the network. Each set of closed firms reduces the total CO$_2$ output by the combined annual CO$_2$ emissions of the respective firms. The closure of firms initializes a shock in the production network which results in the loss of jobs and economic output. These effects are calculated using the ESRI shock propagation algorithm \cite{diem_quantifying_2022} once in the output-weighted and once in the employment-weighted version. The removal of all 119 Hungarian ETS firms results in 31.7\% of job and 38.2\% of economic output loss. This is the same for all decarbonization strategies, but the order in which firms are removed from the production network determines the fractions of expected job and output loss on the way to this final value. The shock-propagation is deterministic, which means that the same set of closed firms always leads to the same outcomes in terms of CO$_2$ reduction, job and output losses. The time horizon of our analysis is one year, as we consider the annual emissions of companies and a shock propagation on a production network that is assumed to remain constant. The estimated job and output losses can therefore be considered worst-case estimates that will likely be smaller when applied to the economy in the real world. Employees who lost their jobs would try to find a new employer. Some jobs might in fact be easily transferred between firms or even sectors, while highly specialized jobs might be harder to replace, see for example \cite{mealy_what_2018} \cite{del_rio-chanona_occupational_2021}. We do not consider these effects in the present framework, but project immediate total potential job loss as a consequence of an imposed decarbonization policy, imposed by a hypothetical social planner. In addition, firms that lost a supplier or a buyer would try to establish new supply relations. In the present modeling framework this is only captured heuristically by assuming that firms with low market shares within their respective NACE4 industry sector are more easily replaceable and firms with high market shares are more difficult to replace. Explicitly considering rewiring and the reallocation of jobs during the shock propagation remains future work. In theory, all combinations of removals of the 119 ETS firms would need to be tested to find the truly optimal strategy with respect to maximum CO$_2$ reduction and minimal expected job and output loss. Since this would result in a combinatorial explosion of possibilities, our goal here is to find a satisfying heuristic that allows for acceptable levels of expected job and output loss for a given CO$_2$ reduction target. In total, we test eleven different heuristics for their potential to rapidly reduce CO$_2$ emissions while securing high levels of employment and economic output. The outcomes for the four main decarbonization strategies are displayed in Fig. \ref{fig:cumulative_plots_strategies} and discussed in the subsequent section. The remaining strategies are shown and discussed in the SI section S2.
\\

\subsection*{Data availability}
Data on financial transactions between Hungarian value-added tax paying firms that support the findings of this study are available at the National Bank of Hungary but restrictions apply to the availability of these data, which were used under license for the current study through directly accessing the servers of the National Bank of Hungary, and so are not publicly available. Requests for collaborations can be addressed to: \url{olahzs@mnb.hu}

\subsection*{Code availability}
The Python library pyeutl \cite{jan_abrell_euetsinfo_2022} which was used to obtain the CO$_2$ emissions of Hungarian firms is open-source and available at \url{https://github.com/jabrell/pyeutl}.
The core components of the analysis, i.e. the extraction of CO$_2$ emissions data for Hungarian ETS firms and the initialization of the studied decarbonization strategies are available at: \url{https://github.com/jo-stangl/reducing_employment_and_economic_output_loss_in_rapid_decarbonization_scenario}

\subsection*{Acknowledgements} 
This work was supported in part by the Austrian Federal Ministry for Climate Action, Environment, Energy, Mobility, Innovation and Technology as part of the funding project GZ 2021-0.664.668 (S.T.), the Austrian Science Fund FWF under P 33751 (S.T.), the Austrian Science Promotion Agency, FFG project under 39071248 (S.T.) and the OeNB Hochschuljubil\"aumsfund P18696 (S.T.). We thank Janos Kertesz for helpful discussions.

\subsection*{Author Contributions Statement}
J.S. and S.T. conceived the work. A.B. cleaned and prepared the data. J.S. and A.B. wrote the code. J.S. A.B., C.D., T.R., S.T. performed the data analysis. All authors analyzed and interpreted the results. J.S. and S.T. wrote the paper. All contributed towards the final manuscript.

\subsection*{Competing Interests Statement}
The authors declare no competing interests.

\FloatBarrier

\bibliography{main}

\FloatBarrier

\onecolumngrid

\newpage
\section*{Supplementary Information}

% \subsection*{Firm-level supply chains to minimize decarbonization unemployment and economic losses}

\subsection*{S1 Short-term vs. long-term models for decarbonization}

Much of the literature on decarbonization strategies and mitigation pathways focuses on mid- to long-term technological change, such as the transition from fossil-based energy provision to renewable sources of energy. Models based on technological change usually center on the efficiency and cost-effectiveness to identify cost-optimal greenhouse gas mitigation pathways for a particular economic unit: a firm, an economic sector, or a national economy. The concept of Marginal Abatement Costs (MAC) is widely used to assess the cost-effectiveness of mitigation options. MACs represent the costs of reducing additional units of pollution, such as CO$_2$ emissions. The abatement of higher levels of CO$_2$ emissions typically comes with increasingly higher costs which are displayed in Marginal Abatement Cost Curves (MACCs). Studies on MACCs focus either on firms \cite{baudry_technological_2021}, economic sectors \cite{gambhir_reducing_2015}, or national economies \cite{wachter_usefulness_2013}. Different methods are used to determine MACs, either bottom-up, that explicitly consider technological mitigation options and their individual costs, or top-down, that derive overall economic costs for different levels of mitigation, often in a computable general equilibrium (CGE) setting \cite{huang_applicability_2016,jiang_hotspots_2020}. The present study presents a complementary perspective to these works. We explicitly focus on the short-term to analyze the role of the firm-level production network in the propagation and amplification of economic shocks originating from rapid decarbonization.

\subsection*{S2 Additional decarbonization strategies}

\begin{figure*}[h!]
\centering
\includegraphics[width=12cm]{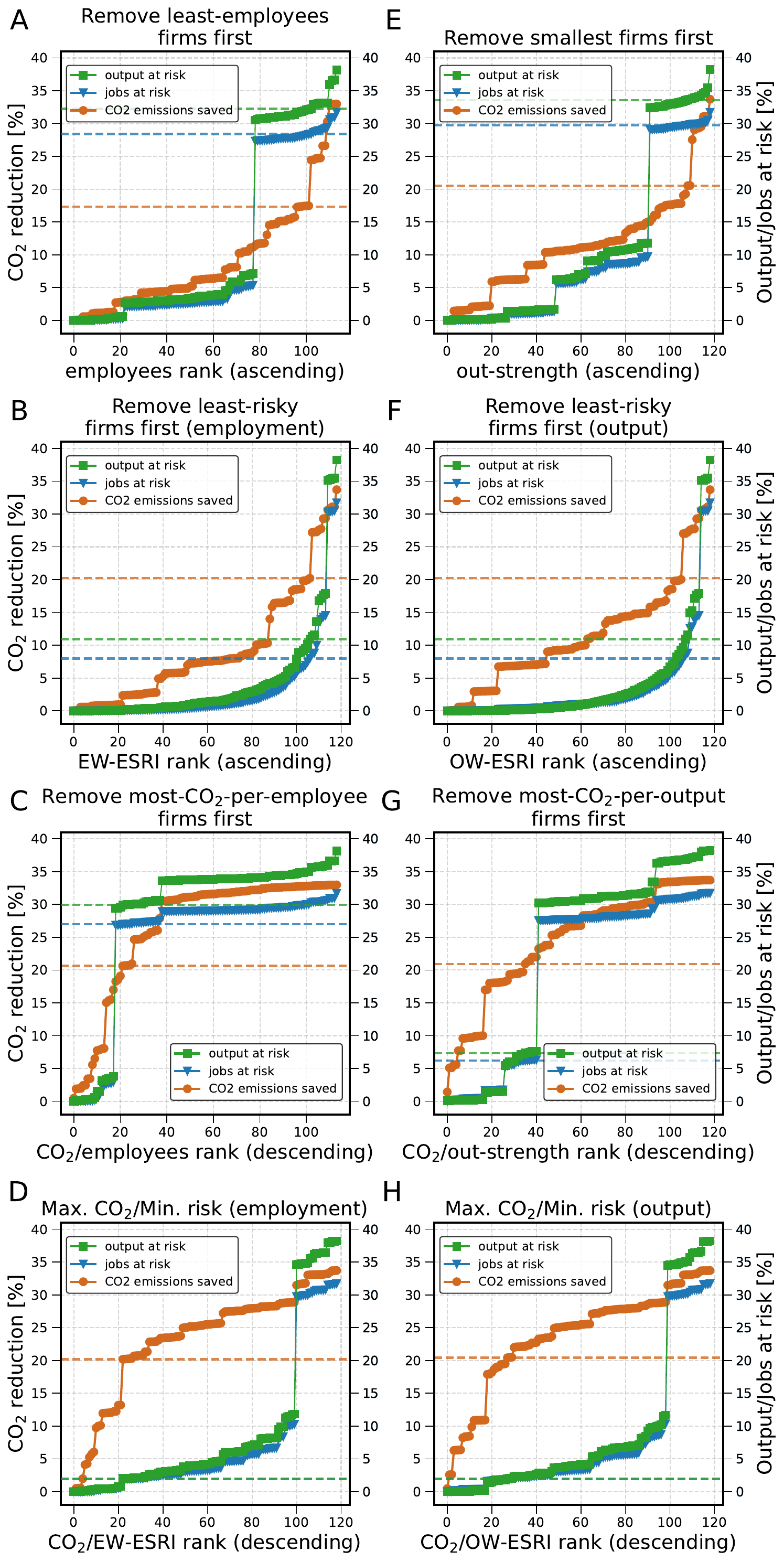}
\caption{Decarbonization strategies centering either around minimal job loss (left column) or minimal loss of economic output (right column).}

\label{fig:remaining_strategies}
\end{figure*}

Figure \ref{fig:remaining_strategies} displays eight different decarbonization strategies. The left column of figures (A),(B),(C), and (D) corresponds to heuristics centered around minimal job loss. The right column of figures (E),(F),(G), and (H) correspond to heuristics centered around minimal economic output loss. The strategies (A) `Remove least-employees firms first', (B) `Remove least-risky firms first (employment)', and (D) `Max. CO$_2$/Min. risk (employment)' are featured in Fig. 3 the main text. Strategy (D) is called `Smart strategy' there. In general, both figure columns display similar characteristics within each row. The heuristics (A) number of employees per firm and (E) out-strength per firm both fail to anticipate the effects of the closure of a systemically very important firm. (B) EW-ESRI and (F) OW-ESRI secure high levels of employment and economic output, but fail to identify decarbonization leverage points. The heuristic (C) CO$_2$/employees delivers significantly worse outcomes than heuristic (G) CO$_2$/out-strength. But only heuristics (D) CO$_2$/EW-ESRI and (H) CO$_2$/OW-ESRI manage to significantly reduce emissions while securing high levels of employment and economic output. The outcomes of the latter are very similar, due to the high correlation between EW-ESRI and OW-ESRI, as discussed in section S6 OW-ESRI vs. EW-ESRI. Due to the high similarity of the results for both sets of strategies, focusing either on employment or output-related heuristics, we decided to only include employment-centered strategies in the main text. \\

\begin{figure}[tb]
\centering
\includegraphics[width=16cm]{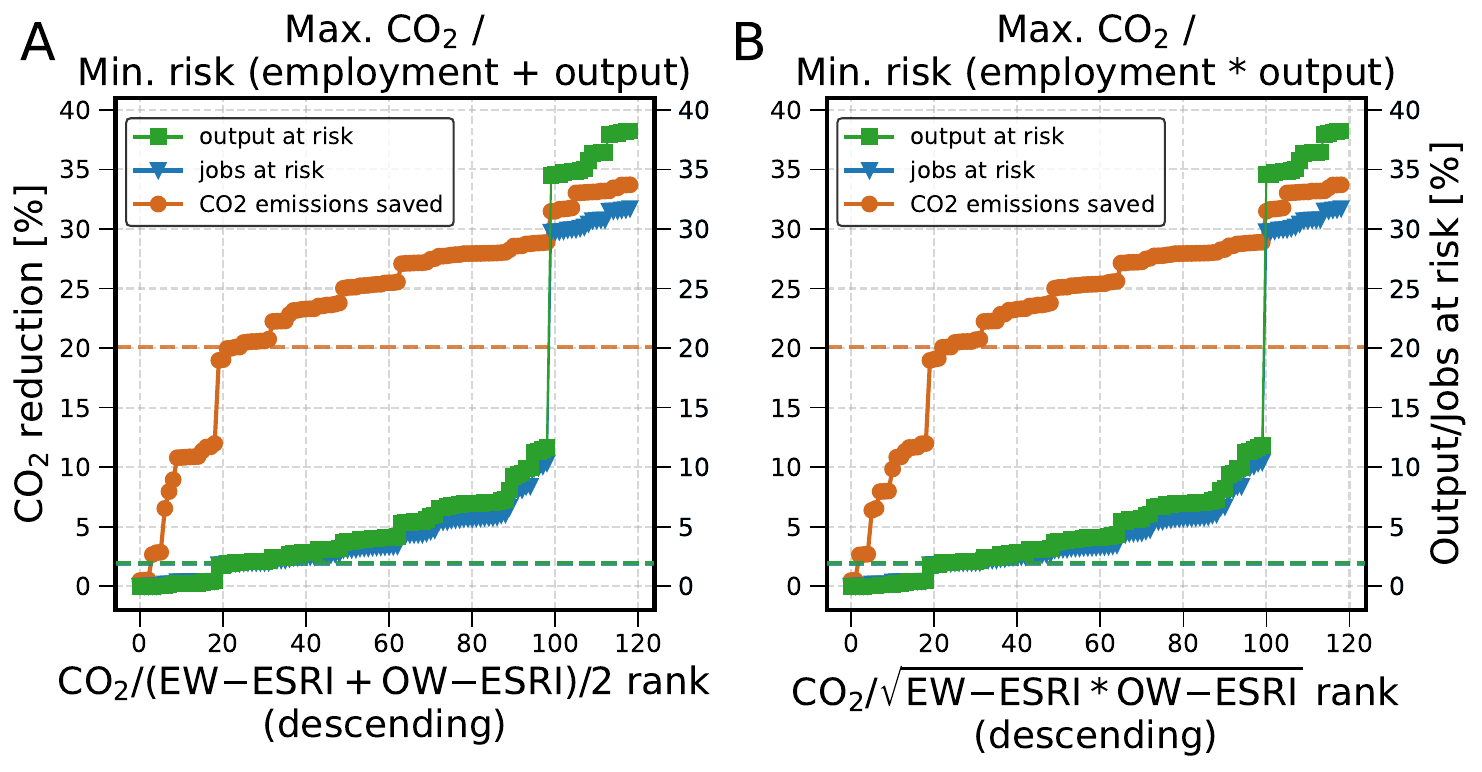}
\caption{Decarbonization strategies incorporating the information captured by both the EW-ESRI and the OW-ESRI. In (A) CO$_2$ emissions are divided by the arithmetic mean of EW-ESRI and OW-ESRI. In (B) CO$_2$ emissions are divided by the geometric mean of EW-ESRI and OW-ESRI.}

\label{fig:arithmetic_geometric_mean}
\end{figure}

Figure \ref{fig:arithmetic_geometric_mean} displays two strategies that aim at incorporating the information about the production network captured by both EW-ESRI and the OW-ESRI. In the strategy `Max. CO$_2$/Min. risk (employment + output)' presented in Fig. \ref{fig:arithmetic_geometric_mean}A CO$_2$ emissions are divided by the arithmetic mean of EW-ESRI and OW-ESRI. In the strategy `Max. CO$_2$/Min. risk (employment * output)' presented in Fig. \ref{fig:arithmetic_geometric_mean}B CO$_2$ emissions are divided by the geometric mean of EW-ESRI and OW-ESRI. Both strategies exhibit similarly good outcomes in terms of CO$_2$ reduction, high levels of remaining employment, and economic output. Table \ref{tab:all_strategies} shows that these strategies do not yield significantly better outcomes than the strategies in which CO$_2$ emissions are divided by either EW-ESRI or OW-ESRI alone. This is why we decided to put them in the SI. 

\FloatBarrier

\FloatBarrier

\begin{sidewaystable}
\begin{tabular}{lrrrrrrrrr}
\toprule
                                   strategy & \makecell{CO$_2$\\saved {[}\%{]}} & \makecell{job loss\\ {[}\%{]}}& \makecell{ output loss\\{[}\%{]}} & \makecell{number of \\firms removed} & \makecell{A -\\Agriculture} & \makecell{B - Mining \\and quarring} & \makecell{C -\\Manufacturing} & \makecell{D -\\Power} & \makecell{H -\\Transportation\\and storage} \\

\midrule
              Remove largest emitters first &       20.252 &  28.559 &  32.608 &                       7 &               0 &                       2 &                 1 &         2 &                              1 \\
         Remove least-employees firms first &       17.465 &  28.353 &  32.155 &                     102 &               1 &                       2 &                55 &        41 &                              3 \\
                Remove smallest firms first &       20.543 &  29.728 &  33.587 &                     109 &               1 &                       3 &                55 &        46 &                              3 \\
Remove least-risky firms first (employment) &       20.211 &   7.973 &  10.934 &                     107 &               1 &                       3 &                56 &        44 &                              2 \\
    Remove least-risky firms first (output) &       20.026 &   7.296 &   9.233 &                     106 &               1 &                       3 &                55 &        44 &                              2 \\
Remove most-CO$_2$-per-employee firms first &       20.632 &  26.986 &  29.932 &                      22 &               0 &                       0 &                 8 &        13 &                              0 \\
  Remove most-CO$_2$-per-output firms first &       20.928 &   6.184 &   7.341 &                      36 &               1 &                       1 &                12 &        19 &                              2 \\
         Max. CO$_2$/Min. risk (employment) &       20.186 &   1.916 &   2.019 &                      23 &               0 &                       1 &                 2 &        17 &                              2 \\
             Max. CO$_2$/Min. risk (output) &       20.421 &   1.937 &   2.003 &                      28 &               0 &                       1 &                 5 &        19 &                              2 \\
Max. CO$_2$/Min. risk (employment + output) &       20.079 &   1.910 &   1.976 &                      24 &               0 &                       1 &                 2 &        18 &                              2 \\
Max. CO$_2$/Min. risk (employment * output) &       20.057 &   1.906 &   1.971 &                      23 &               0 &                       1 &                 1 &        18 &                              2 \\
\bottomrule
\end{tabular}

\caption{Results for the 20\% emission reduction benchmark for all studied decarbonization strategies}
\label{tab:all_strategies}
\end{sidewaystable}

\FloatBarrier

\subsection*{S3 Accounting for network effects lead to significantly higher estimates of job and output loss}
\begin{figure*}[h!]
\centering
\includegraphics[width=16cm]{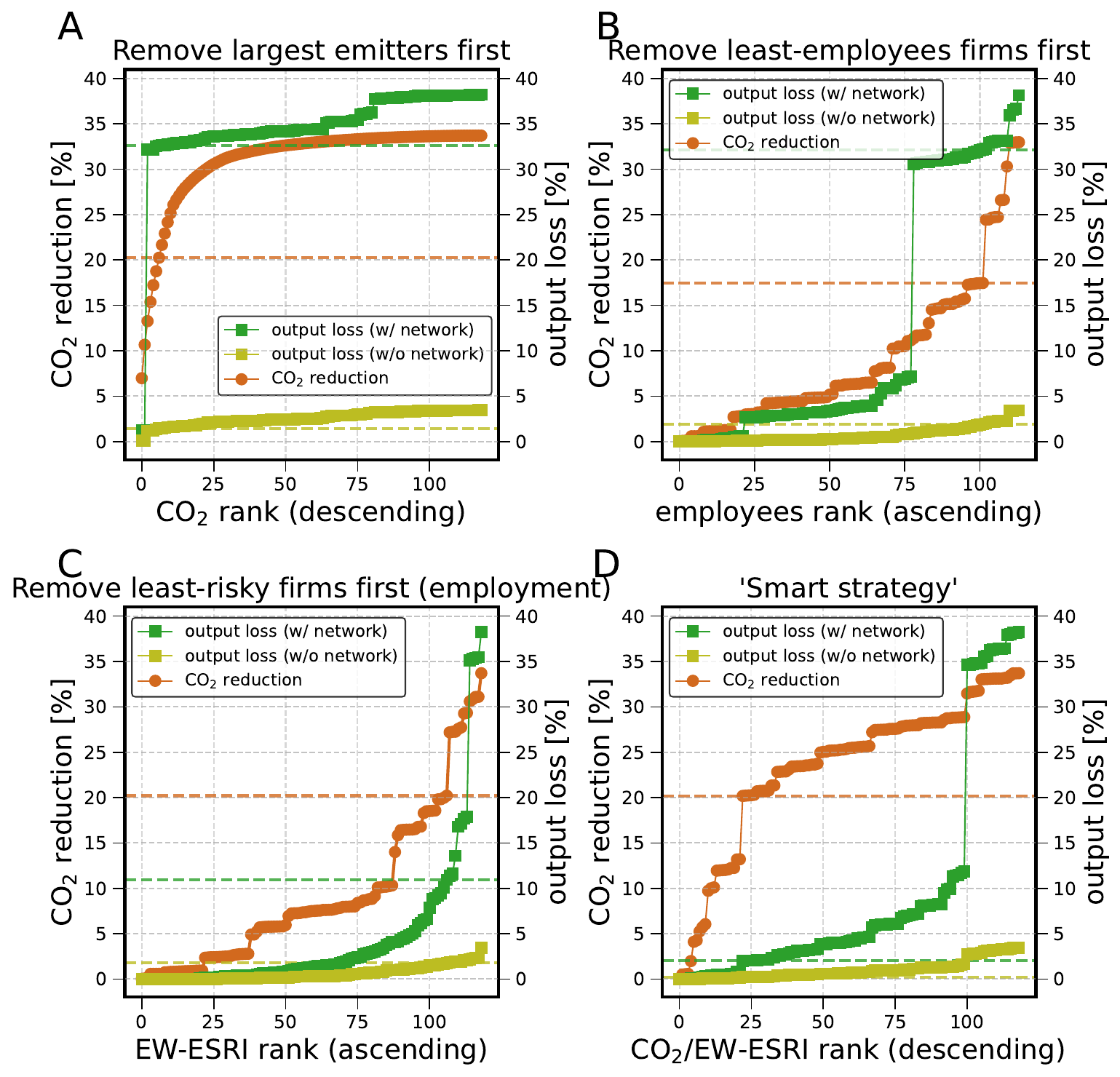}
\caption{ Economic output loss with and without accounting for network effects in selected decarbonization scenarios.}

\label{fig:linear_vs_network_effect_OUTSTRENGTH}
\end{figure*}

\begin{figure*}[h!]
\centering
\includegraphics[width=16cm]{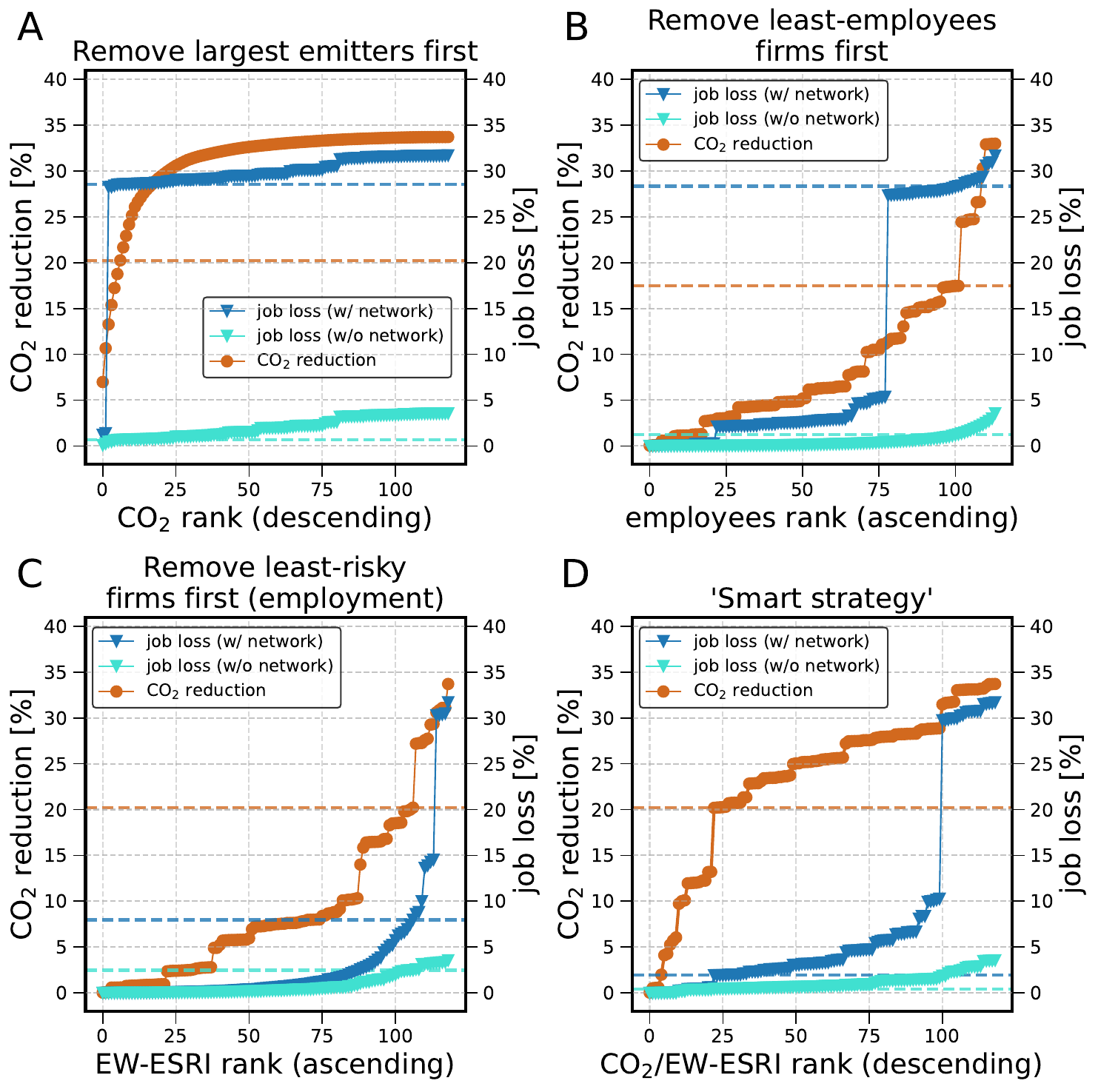}
\caption{Job loss with and without accounting for network effects in selected decarbonization scenarios-}

\label{fig:linear_vs_network_effect_EMPLOYEES}
\end{figure*}

To estimate the total potential job and economic output loss in our hypothetical decarbonization strategies, we again use the EW-ESRI and the OW-ESRI. Using these measures allows us to take the indirect effects in the production network into account. In our framework, if a firm is removed from the production network, it causes a shock that propagates upstream to its suppliers and downstream to its customers. Depending on their specific production functions, the customers of the removed firm have to reduce their output which also leads to a reduction of its workforce. Likewise, the suppliers of the removed firm now face decreased demand for their products which leads to a reduction of their output and their workforce. These shocks propagate through the production network until a new equilibrium state is reached. Taking the network the targeted firms are embedded in into account therefore leads to significantly higher estimates of jobs and economic output affected for each decarbonization scenario. Figure \ref{fig:linear_vs_network_effect_OUTSTRENGTH} depicts the four decarbonization scenarios from Fig. 3 in the main text. Economic output loss is once estimated taking the network into account, i.e. by sequentially calculating the OW-ESRI of removing the Hungarian ETS firms according to the respective decarbonization strategy (dark green). The second estimate does not consider the network dynamics and is derived by taking the sum of the output losses of the directly removed firms divided by the total output of all firms in the network (light green). Each sub-figure of Fig. \ref{fig:linear_vs_network_effect_OUTSTRENGTH} shows a significant difference in estimated output loss when considering network effects compared to when not considering them. \\
Likewise Fig. \ref{fig:linear_vs_network_effect_EMPLOYEES} depicts the four decarbonization scenarios from Fig. 3 in the main text with regards to job loss. Job loss is once estimated taking the network into account, i. e. by sequentially calculating the EW-ESRI of removing the Hungarian ETS firms according to the respective decarbonization strategy (dark blue). The second estimate does not consider the network dynamics and is derived by summing up the number of jobs that would be lost from the directly removed firms divided by the total number of observed jobs (= 2.333.975, light blue). Again, each sub-figure of Fig. \ref{fig:linear_vs_network_effect_OUTSTRENGTH} shows a significant difference in job loss when considering the network compared to when not considering it. \\
In Tab. \ref{tab:linear_vs_network_effects} the results for the CO$_2$ reduction benchmark of -20 $\%$ are shown for both the 'direct' estimates of job and economic output losses and the respective estimates taking the network effects into account. Only accounting for the 'direct' effects of removing companies from the production network leads to relatively minor estimates of job and economic output losses in all decarbonization scenarios. Systemic events that are triggered by removing highly important firms from the production network cannot be captured by such a simple accounting framework. Using the EW-ESRI and the OW-ESRI that explicitly consider these network effects leads to 3 to 42 times higher estimates of job
loss and 6 to 23 times higher estimates of economic output loss for the described scenarios. In the present framework, in which the network is aggregated to one year and kept static during the shock propagation, these estimates are likely too pessimistic. But it arguably still captures an important part of reality - the dependence of firms on each other through their supply relations.

\FloatBarrier

\begin{table*}[h!]
\makebox[\linewidth]{

\begin{tabular}{lrrrrrrrr}
\toprule
                                    scenario & \makecell{CO$_2$\\reduction [\%]} & \makecell{job loss [\%]\\(w/ network)} & \makecell{job loss [\%]\\(w/o network)} & \makecell{network\\amplification\\job loss} & \makecell{output loss [\%]\\(w/ network)} & \makecell{output loss [\%]\\(w/o network)}& \makecell{network\\amplification\\output loss} & \makecell{firms\\removed} \\
\midrule
              Remove largest\\emitters first &        20.25 &                     28.56 &                       0.68 &                      42.000000 &                        32.61 &                          1.44 &                         22.645833 &                       7 \\
         Remove least-employees\\firms first &        17.46 &                     28.35 &                       1.27 &                      22.322835 &                        32.15 &                          1.90 &                         16.921053 &                     102 \\
Remove least-risky\\firms first (employment) &        20.21 &                      7.97 &                       2.47 &                       3.226721 &                        10.93 &                          1.78 &                          6.140449 &                     107 \\
                           'Smart strategy' &        20.19 &                      1.92 &                       0.44 &                       4.363636 &                         2.02 &                          0.22 &                          9.181818 &                      23 \\
\bottomrule
\end{tabular}
}
\caption{Comparison of the outcomes for the 20$\%$ CO2 emission reduction target for the four decarbonization strategies presented in Fig. \ref{fig:linear_vs_network_effect_OUTSTRENGTH} and Fig. \ref{fig:linear_vs_network_effect_EMPLOYEES}. Percentages for job loss are given relative to the total number of observed jobs in our dataset (= 2333975) and percentages for output loss are given relative to total output of all firms in the production network. Considering network effects leads to drastically higher estimates of job and economic output loss compared to considering only `direct' effects.}
\label{tab:linear_vs_network_effects}
\end{table*}

\FloatBarrier

\subsection*{S4 Reconstructing the Hungarian firm-level production network}
\begin{figure*}[h!]
\centering
\includegraphics[width=16cm]{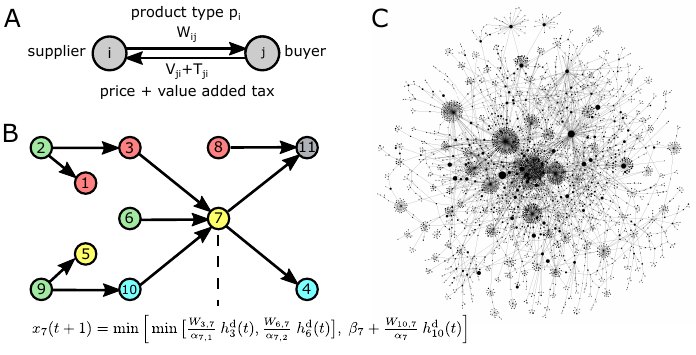}
\caption{Reconstruction of the firm-level production network of Hungary via value-added tax (VAT) data. (A) Schematic supplier-buyer relationship between two firms, $i$ and $j$. Supplier $i$ delivers the amount (monetary value) $W_{ij}$ of product $p$ to buyer $j$. The buyer $j$ pays the product price $V_{ji}$ plus the value-added tax $T_{ji}$. (B) Schematic production sub-network consisting of 11 firms. Color represents industry affiliation (NACE). The elements of the weighted adjacency matrix $W_{ij}$ are the transacted product volumes in monetary values. $h_i^d$ represents the relative production level of firm $i$ at time $t$. The production function of firm 7 is shown as an example. (C) A small subset of the Hungarian production network with 4,070 firms and 4,845 supply relations. Node size corresponds to total strength and is a proxy for firm size. For more details on the reconstruction of the network, see \cite{diem_quantifying_2022} and its Supplementary Information.
}

\label{Fig.:reconstruction}
\end{figure*}

The empirical foundation of the present study is the reconstructed firm-level production network of the Hungarian economy. Since 2014, the Hungarian National Tax and Customs Administration has been collecting value-added tax (VAT) transactions between all VAT-tax-paying Hungarian firms. This data is available through the Central Bank of Hungary. From this data, the supply relations between firms can be inferred, which allows us to create network snapshots of the practically complete Hungarian firm-level production network at a given time. In principle, the ideal way to assess the input-output structure of firms would be to use physical quantities. To our knowledge, comprehensive data of this kind is not available for an entire national economy. We therefore follow the approach of Diem et al. \cite{diem_quantifying_2022} and use monetary flows between firms as proxies to characterize the supply relations and the production functions of firms. As this approach has already been used by several studies on economic shock spreading \cite{inoue_firm-level_2019}\cite{carvalho_supply_2021}, we think it is also justified in the present study. It is important to note that we use a static representation of the Hungarian production network and aggregate supply relation between firms for the time period of one year. Figure \ref{Fig.:reconstruction} schematically shows how the Hungarian firm-level production network is obtained for a given year. In Fig. \ref{Fig.:reconstruction}a the supplier-buyer relationship of two representative firms is depicted. Supplying firm $i$ delivers a monetary value of $W_{ij}$ of product $p$ to the buying firm, $j$. Buying firm $j$ pays price $V_{ji}$ and the value-added tax, $T_{ji}$. Since $T_{ji}$ is known, $W_{ij}$ can be inferred.  Figure \ref{Fig.:reconstruction}B shows a small subset of the production network. Every firm is affiliated with an industry sector (color) that determines the (non-)`essentialness' of its products for firms of every other industry sector. Each firm is also assigned a generalized Leontief production function introduced in \cite{diem_quantifying_2022}. It determines how a firm's inputs (from other firms) are converted into outputs. Here, the generalized Leontief production function for firm 7 is shown as an example.

\begin{equation}
    x_7(t+1) = \text{min } \bigl[ \text{min } [\frac{W_{3,7}}{\alpha_{7,1}}\text{ } h_3^d(t), \text{ } \frac{W_{6,7}}{\alpha_{7,2}} \text{ } h_6^d(t) ] ,\text{ } \beta_7 + \frac{W_{10,7}}{\alpha_{7}} \text{ } h_{10}^d(t) \bigr]
\end{equation}

\noindent Firm 7 has two essential inputs from firms 3 and 6 and one non-essential input from firm 10. Its output $x_7$ at time $t+1$ is determined by the relative downstream constrained production levels, $h_i^d = \frac{x_i^d(t)}{x_i(0)}$, of firms 3, 6 and 7 at time step $t$. $x_i^d(t)$ is the absolute downstream constrained production level of firm $i$ at time $t$ and $x_i^u(t)$ is the upstream constrained production level of firm $i$ at time $t$. This distinction allows for the separate treatment of supply shocks (passed on downstream) and demand shocks (passed on upstream). The relative upstream constrained production level of firm $i$ at time $t$ is defined analogously as, $h_i^u(t) = \frac{x_i^u(t)}{x_i(0)}$. $\alpha_{ik}$ are technologically determined coefficients and $\beta_i$ is the production level of firm $i$ that is possible without non-essential inputs. $\alpha$ and $\beta$ are determined by $W_{ij}$ and the set of essential inputs of firm $i$, $\mathcal{I}_i^{es}$ and its set of non-essential inputs, $\mathcal{I}_i^{ne}$. For more details, see \cite{diem_quantifying_2022} and its Methods section. Panel \ref{Fig.:reconstruction}C shows a subset of the Hungarian firm-level production network, containing 4,070 firms and 4,845 supply relations. It is visible that the supply chains of firms are highly interconnected, a fact that gives rise to a complex production network consisting of many hubs and cycles.

\subsubsection*{Value-added tax (VAT) reporting in Hungary}
From the beginning of 2015 to the second quarter of 2018, only transactions exceeding 1 million HUF of tax content for the sum of the transactions between two firms in a given reporting period (monthly, quarterly, annually) were recorded. In the period from the third quarter of 2018 to the second quarter of 2020, the threshold was lowered to 100,000 Hungarian Forint (HUF), but the threshold now referred to individual transactions. This brought many more firms and supply relations into view, while some firms dropped out of the dataset when their individual transactions stayed below this new threshold. As of the third quarter of 2020, there is no longer a threshold and all inter-firm invoices must be reported. For this study we focus on the year 2019 for two reasons: first, at the time of conducting our analysis, 2019 was the most recent year for which CO$_2$ emissions data has been available through the EU Emission Trading System (ETS). Second, 2019 was the last year before the COVID-19 pandemic and the subsequent economic crisis, which disrupted many supply relations between firms and arguably makes the network snapshots of these subsequent years unrepresentative as initial states of the Hungarian economy. 

\FloatBarrier

\begin{figure*}[h!]
\centering
\includegraphics[width=16cm]{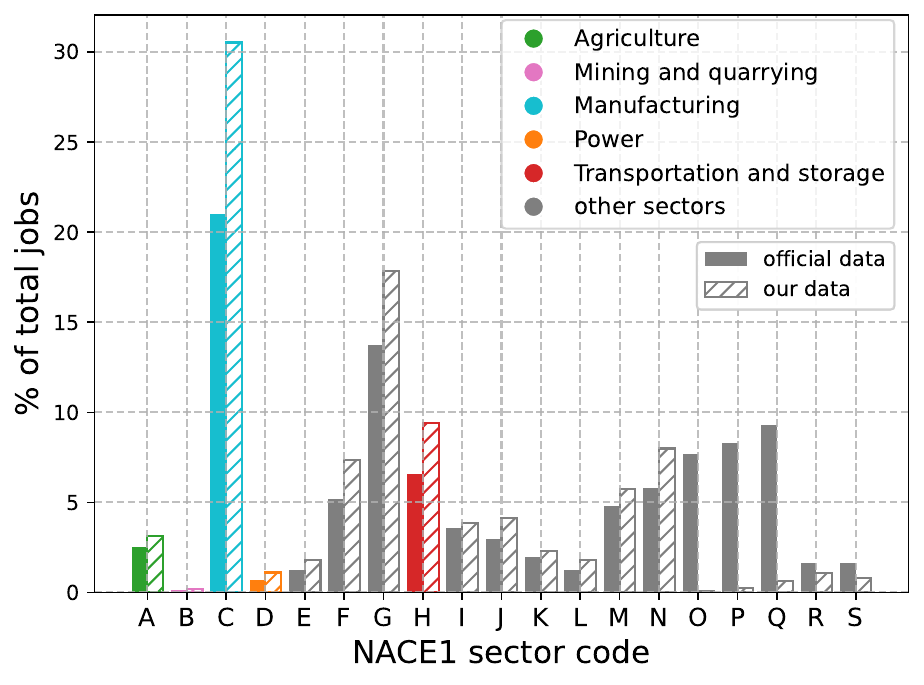}
\caption{ Relative employment in Hungary by NACE level 1 sector according to official statistics \cite{hungarian_central_statistical_office_20225_nodate} and our aggregated firm-level employment data.}

\label{fig:employment_data}
\end{figure*}

\subsection*{S5 EW-ESRI vs. number of employees}
The employment-weighted economic systemic risk index (EW-ESRI) is a measure of potential job loss in case of a production stop of a single company or a set of companies in the entire economy. As explained in the Methods section in the main text, it captures the indirect job losses of firms in the production network whose production becomes constrained due to downstream or upstream shocks caused by the initial set of failing companies.  Empirically, we are able to do this re-weighting of the ESRI for the Hungarian production network since a dataset on the number of employees of a large subset of firms is available to us through the Hungarian Central Bank. This dataset covers 2,333,975 jobs in total, which is 65.6\% of the 3,557,700 employees in the year 2019 according to official statistics \cite{hungarian_central_statistical_office_20225_nodate}. In order to understand the coverage of our firm-level employment data set, we aggregate jobs by NACE level 1 category of their respective employing firms. The coverage of different economic sectors by our employment data varies. In Fig. \ref{fig:employment_data} we compare the two relative distributions of employees per Nace level 1 sector of both the official statistics and our aggregated data. The sectors in which Hungarian firms operate that are covered by the EU emission trading system (ETS) are colored, and the remaining sectors are held in gray. As can be seen in Fig. \ref{fig:employment_data}, our firm-level employment data recovers essential features of the official employment distribution, such as the relative importance between sectors. Also, the majority of jobs are located in the manufacturing sector, even though this sector exhibits a higher share of 30.53\% of employees in our data set. In general, our data set exhibits the best coverage in the producing and in the service sectors and shows the worst coverage in the sectors higher up in the NACE classification scheme, like the public or education sector.

\begin{figure*}[h!]
\centering
\includegraphics[width=12cm]{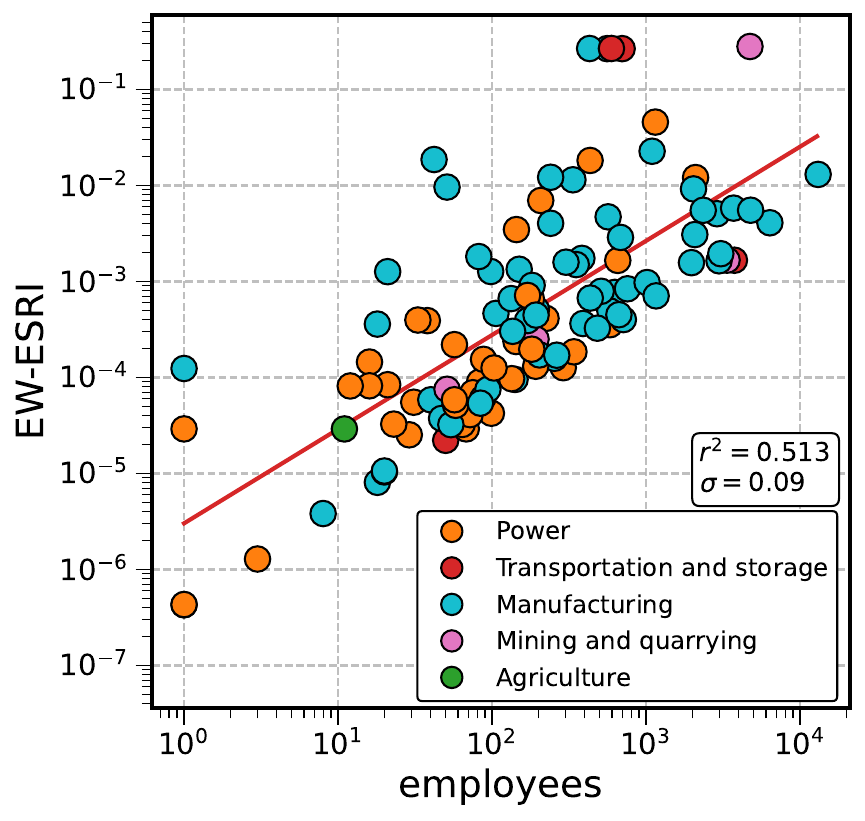}
\caption{EW-ESRI vs. number of employees.}

\label{fig:ew-esri_vs_employees}
\end{figure*}

Figure \ref{fig:ew-esri_vs_employees} depicts the relationship between the EW-ESRI and the number of employees for each company in the Hungarian ETS in the year 2019. We observe that the logarithms of EW-ESRI and the number of employees are correlated with an $r^2 = 0.513$ and a standard deviation $\sigma = 0.09$. This means that for a fair share of the sample the EW-ESRI and the number of employees per firm are correlated, but for some firms, the two measures deviate quite a lot. This is especially true for firms with high systemic relevance as observed in the upper right corner of Fig. \ref{fig:ew-esri_vs_employees}. Note the imaginary line bounding the EW-ESRI from below. This lower bound is due to the fact that the employment impact of a firm on the network cannot be smaller than its own number of employees. Also note three firms with supposedly only one employee in the lower left corner of Fig. \ref{fig:ew-esri_vs_employees}. According to our data, these firms were in the process of being closed in the year 2019 which explains their curious employment numbers.

Table \ref{tab:employment} displays the coverage of employment data for each NACE level 1 sector in Hungary. Figure \ref{fig:employment_data} visualizes the relative distributions of jobs per NACE1 sector for both the official statistics and our data. In general the data matches remarkably well in absolute terms for most economic sectors, especially for producing sectors. Service sectors like public administration and education are worst covered by our data. Note that for sector `D - Electricity, Gas, Steam And Air Conditioning Supply' the number of employees in our data slightly exceeds the number of employees in the official statistics. This might look puzzling at first sight, but is explained by a mismatch of NACE classification between the official statistics and our firm-level production network dataset. We see some firms belonging to NACE level 1 sector D that are probably differently classified at the Hungarian statistical office which leads to such an overestimation. The official data stems from the Hungarian statistical office \cite{hungarian_central_statistical_office_20225_nodate}.

\begin{table*}[p]
\centering
\begin{tabular}{llrrrrr}
\toprule
\makecell{NACE1\\code} &                                  \makecell{NACE1\\description} & \makecell{employees\\in official data} & \makecell{employees\\in our data}  & \makecell{relative share\\in official data} & \makecell{relative share\\in our data} \\
\midrule
         A &                  Agriculture, Forestry And Fishing &                     89000 &                       73662 &                            2.50 &                       3.16 \\
         B &                               Mining And Quarrying &                      4200 &                        4094 &                            0.12 &                       0.18 \\
         C &                                      Manufacturing &                    747000 &                      712539 &                           21.00 &                      30.53 \\
         D & Electricity, Gas, Steam And Air Conditioning Su... &                     24600 &                       25891 &                            0.69 &                       1.11 \\
         E & Water Supply; Sewerage, Waste Management And Re... &                     44800 &                       42517 &                            1.26 &                       1.82 \\
         F &                                       Construction &                    185500 &                      171237 &                            5.21 &                       7.34 \\
         G & Wholesale And Retail Trade; Repair Of Motor Veh... &                    488900 &                      416039 &                           13.74 &                      17.83 \\
         H &                         Transportation And Storage &                    233100 &                      219907 &                            6.55 &                       9.42 \\
         I &          Accommodation And Food Service Activities &                    127900 &                       89907 &                            3.60 &                       3.85 \\
         J &                      Information And Communication &                    104800 &                       96068 &                            2.95 &                       4.12 \\
         K &                 Financial And Insurance Activities &                     70000 &                       53325 &                            1.97 &                       2.28 \\
         L &                             Real Estate Activities &                     44200 &                       42478 &                            1.24 &                       1.82 \\
         M &  Professional, Scientific And Technical Activities &                    169900 &                      134320 &                            4.78 &                       5.75 \\
         N &      Administrative And Support Service Activities &                    207200 &                      186534 &                            5.82 &                       7.99 \\
         O & Public Administration And Defence; Compulsory S... &                    273700 &                        1526 &                            7.69 &                       0.07 \\
         P &                                          Education &                    295900 &                        5310 &                            8.32 &                       0.23 \\
         Q &            Human Health And Social Work Activities &                    330500 &                       14751 &                            9.29 &                       0.63 \\
         R &                 Arts, Entertainment And Recreation &                     59000 &                       24946 &                            1.66 &                       1.07 \\
         S &                           Other Service Activities &                     57500 &                       18924 &                            1.62 &                       0.81 \\
\bottomrule
 & SUM & 3557700 & 2333975 & 100.00 & 100.00
\end{tabular}

\caption{Employment in each NACE1 sector in official employment statistics \cite{hungarian_central_statistical_office_20225_nodate} and our firm-level employment data set}
\label{tab:employment}
\end{table*}

\FloatBarrier

\subsection*{ S6 OW-ESRI vs. out-strength}
\begin{figure*}[h!]
\centering
\includegraphics[width=12cm]{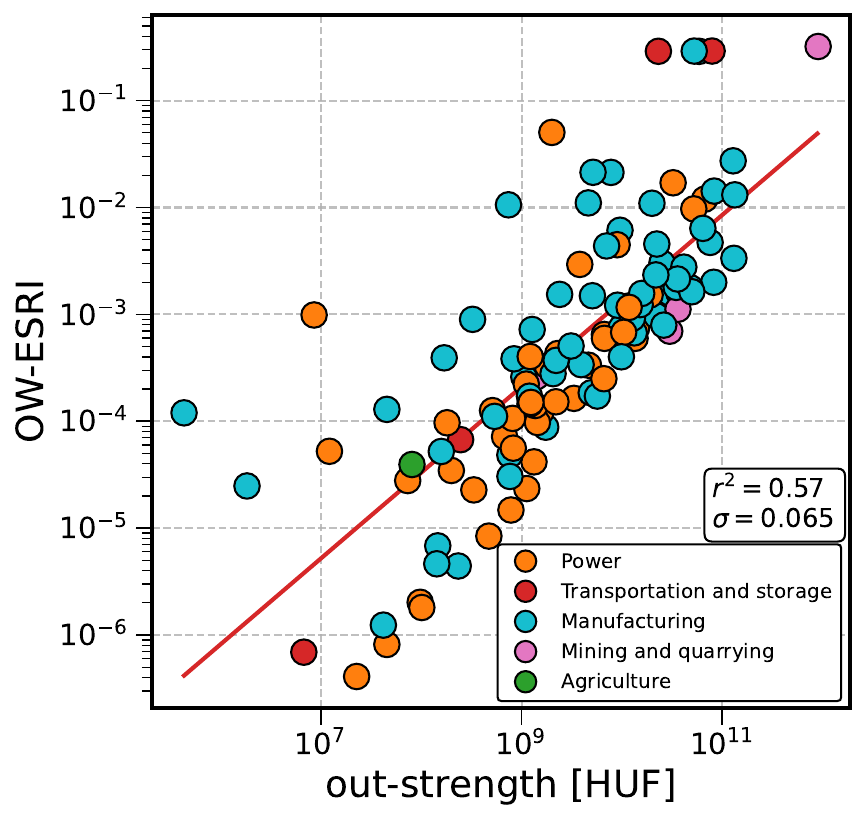}
\caption{OW-ESRI vs. out-strength.}

\label{fig:ow-esri_vs_out-strength}
\end{figure*}

The output-weighted economic systemic risk index (OW-ESRI) is a measure of potential economic output loss in case of a production stop of a single company or a set of companies. As explained in the Methods section in the main text, it captures the indirect output losses of firms in the production network whose production becomes constrained due to downstream or upstream shocks caused by the initial set of failing companies. Figure \ref{fig:ow-esri_vs_out-strength} depicts the relationship between the OW-ESRI and out-strength for each company in the Hungarian ETS in the year 2019. We observe that the logarithms of OW-ESRI and out-strength are correlated with an $r^2 = 0.57$ and a standard deviation $\sigma = 0.065$. This means that for a fair share of the sample the OW-ESRI and the out-strength per firm are correlated, but for some firms, the two measures deviate quite a lot. This is especially true for firms with high systemic relevance as observed in the upper right corner of Fig. \ref{fig:ow-esri_vs_out-strength}. Also, some firms with low out-strength display disproportionally high OW-ESRI values. Note again the imaginary line bounding the OW-ESRI from below. This lower bound is due to the fact that the output impact of a firm on the network cannot be smaller than its own out-strength.

\subsection*{S6 Employees vs. out-strength}
\begin{figure*}[h!]
\centering
\includegraphics[width=12cm]{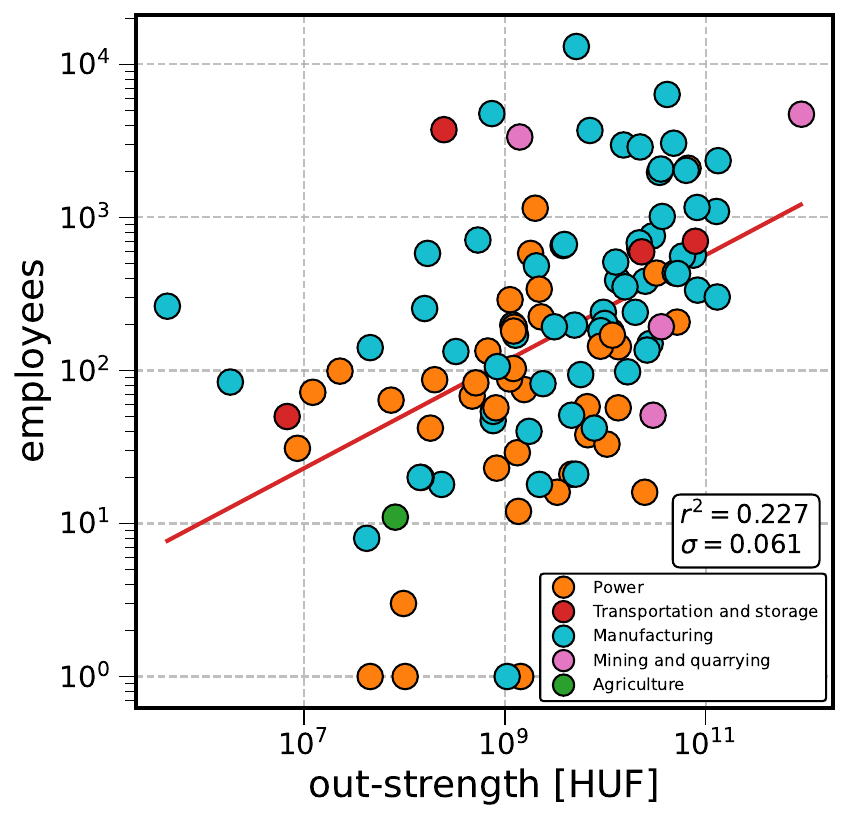}
\caption{Number of employees vs. out-strength}

\label{fig:employees_vs_out-strength}
\end{figure*}

Figure \ref{fig:employees_vs_out-strength} displays number of employees vs. out-strength for each firm in the Hungarian ETS. Even in the double-logarithmic plot the values are quite uncorrelated, exhibiting only an $r^2 = 0.227$. This motivates the separate analysis of decarbonization strategies that focus either on employment or total economic output, as displayed in Fig. \ref{fig:remaining_strategies}. Note the four firms with only one employee, but significant out-strength. As mentioned above, these firms are in the process of being closed according to our data. Interestingly, they still exhibit significant out-strength in the year 2019.

\FloatBarrier

\subsection*{S8 OW-ESRI vs. EW-ESRI}
It is easily visible from the figures comparing decarbonization strategies that OW-ESRI and EW-ESRI are highly correlated. This is due to the fact that they both use the same shock-spreading mechanism. The resulting relative production losses from the shock propagation cascade are then weighted by out-strength and number of employees, respectively. This high correlation indicates that the topological features of the network are much more significant for the estimation of systemic risk than individual firm properties like out-strength or the number of employees.

\begin{figure*}[h!]
\centering
\includegraphics[width=12cm]{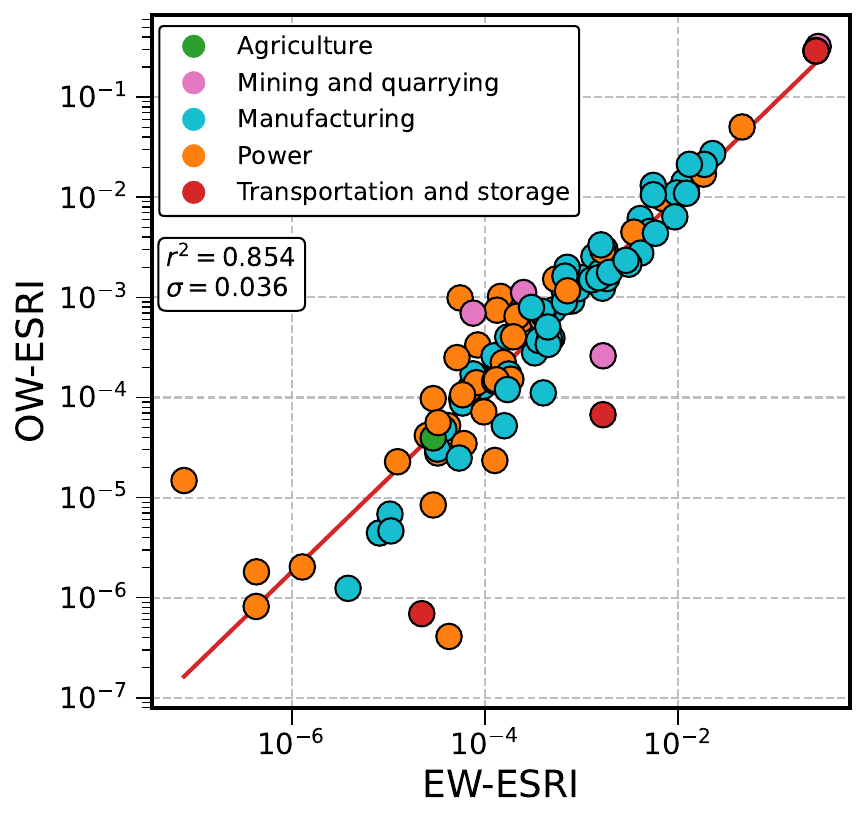}
\caption{OW-ESRI vs. EW-ESRI.}

\label{fig:ow-esri_vs_ew-esri}
\end{figure*}

\FloatBarrier

\subsection*{S9 Market shares of Hungarian companies in the EU Emission Trading System}
\begin{figure*}[h!]
\centering
\includegraphics[width=14cm]{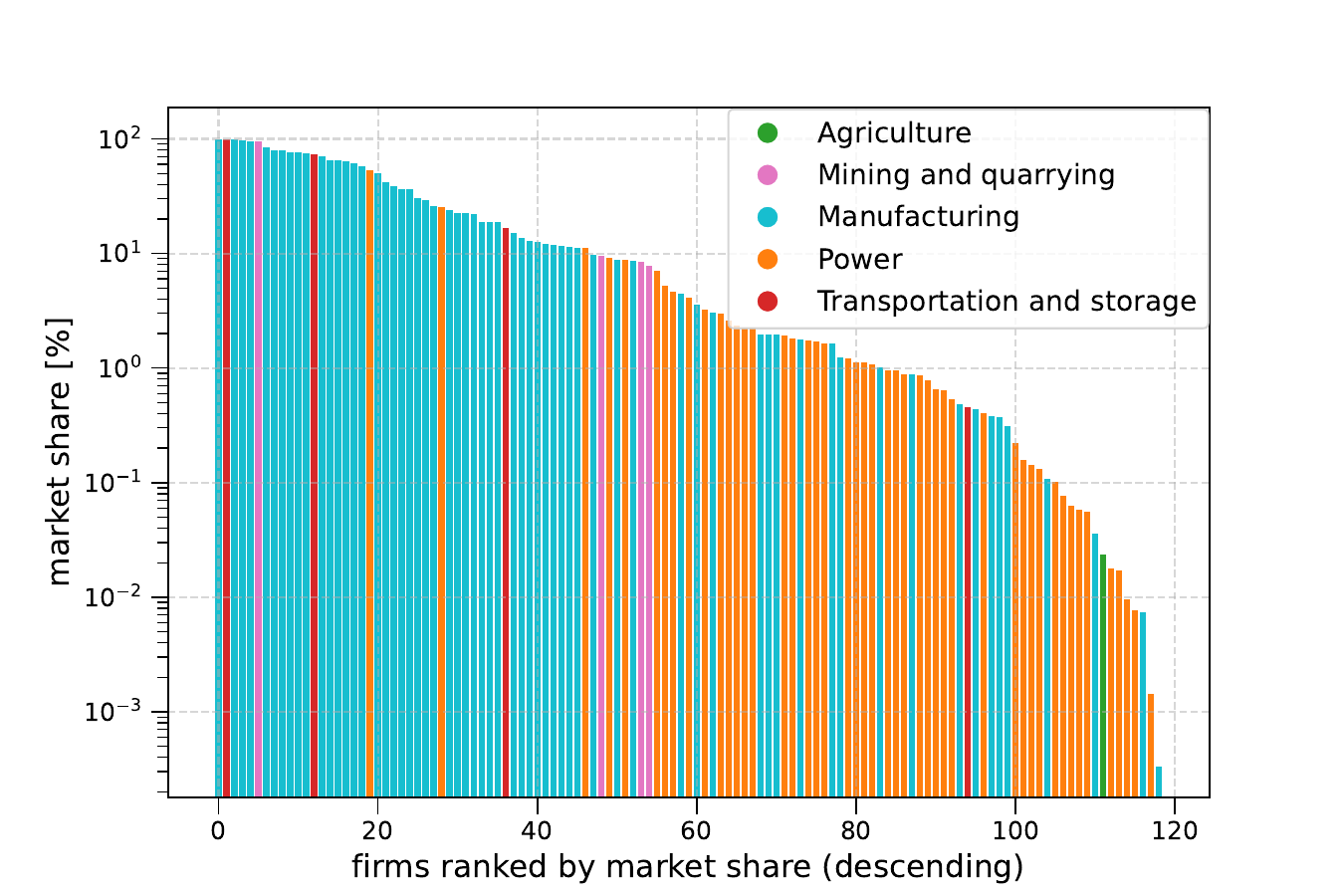}
\caption{Market shares of the ETS firms as given by a firms' out-strength divided by the total out-strength of its NACE2 sector.}

\label{fig:market_shares}
\end{figure*}

The market shares of firms within a given production network are a crucial input to the calculation of the economic systemic risk index (ESRI) and substantially influence the systemic importance of a company within the network \cite{diem_quantifying_2022}. Figure \ref{fig:market_shares} displays the market shares of the ETS companies in Hungary in the year 2019, measured by the out-strength $s_i$ of each firm divided by the total out-strength of the respective NACE level 4 sector. The market shares of companies in the manufacturing sector appear to be higher than the market shares of power companies which explains their bias towards higher systemic risk as e.g. displayed in Fig. 2 in the main text. We know from other sources such as the IEA report ‘Hungary 2022 Energy Policy Review’ that the electricity market in Hungary is quite concentrated in terms of ownership structure \cite{international_energy_agency_hungary_2022}. But the dominant electricity companies consist of many smaller companies which gives each individual electricity firm a much lower market share. We observe these individual companies in our data on the Hungarian production network. Note that some companies exhibit very high market shares of almost 100 percent within their respective NACE4 category. This means that these firms are virtually not replaceable and a shock originating from a removal of those companies will be passed on undiminished.

\FloatBarrier

\FloatBarrier

\end{document}